\documentclass[journal,10pt,twocolumn]{IEEEtran}
\usepackage{xmpaper,lzqsymbol}
\usepackage{amssymb,amsmath,latexsym}
\usepackage{graphicx,subfigure}
\usepackage{color,cite,times}
\usepackage{epstopdf}
\usepackage{multirow}
\usepackage{array}

\usepackage{amsmath}
\usepackage{bm}
\usepackage{algorithm}
\usepackage{algorithmic}
\usepackage[T1]{fontenc}

\usepackage{booktabs}
\usepackage{bbm} 
\usepackage{subfigure}

\begin{document}

\title{Robust Secure UAV Communications with the Aid of Reconfigurable Intelligent Surfaces}
\author{Sixian Li, Bin Duo,~\IEEEmembership{Member,~IEEE}, Marco~Di~Renzo,~\IEEEmembership{Fellow,~IEEE}, Meixia~Tao,~\IEEEmembership{Fellow,~IEEE}, and Xiaojun~Yuan,~\IEEEmembership{Senior~Member,~IEEE}
\thanks{S.~Li, B.~Duo, and X.~Yuan are with the Center for Intelligent Networking and Communications, the National Laboratory of Science and Technology on Communications, the University of Electronic Science and Technology of China, Chengdu 611731,
China (e-mail: sxli@std.uestc.edu.cn; duobin@cdut.edu.cn; xjyuan@uestc.edu.cn). B.~Duo is also with the College of Information Science \& Technology, Chengdu University of Technology, Chengdu 610059, China. 

M. Di Renzo is with Universit\'e Paris-Saclay, CNRS, CentraleSup\'elec, Laboratoire des Signaux et Syst\`emes, 3 Rue Joliot-Curie, 91192 Gif-sur-Yvette, France. (marco.di-renzo@universite-paris-saclay.fr)

M.~Tao is with the Department of Electronic Engineering, Shanghai Jiao Tong University, Shanghai 200240, China (e-mail: mxtao@sjtu.edu.cn).

}

}
\maketitle
\begin{abstract}
    This paper investigates a novel unmanned aerial vehicles (UAVs) secure communication system with the assistance of reconfigurable intelligent surfaces (RISs), where a UAV and a ground user communicate with each other, while an eavesdropper tends to wiretap their information. Due to the limited capacity of UAVs, an RIS is applied to further improve the quality of the secure communication. The time division multiple access (TDMA) protocol is applied for the communications between the UAV and the ground user, namely, the downlink (DL) and the uplink (UL) communications. In particular, the channel state information (CSI) of the eavesdropping channels is assumed to be imperfect. We aim to maximize the average worst-case secrecy rate by the robust joint design of the UAV's trajectory, RIS's passive beamforming, and transmit power of the legitimate transmitters. However, it is challenging to solve the joint UL/DL optimization problem due to its non-convexity. Therefore, we develop an efficient algorithm based on the alternating optimization (AO) technique. Specifically, the formulated problem is divided into three sub-problems, and the successive convex approximation (SCA), $\mathcal{S}$-Procedure, and semidefinite relaxation (SDR) are applied to tackle these non-convex sub-problems. Numerical results demonstrate that the proposed algorithm can considerably improve the average secrecy rate compared with the benchmark algorithms, and also confirm the robustness of the proposed algorithm. 
    
\end{abstract}

\section{Introduction}
With the rapid growth of the number of network devices, it is expected that the overall mobile data traffic will reach astonishingly up to 77 exabytes per month by 2022 \cite{macro2019}, which undoubtedly poses a tremendous challenge for current mobile communication networks. To meet these explosive demands, innovative wireless transmission technologies have been investigated in the past few years, such as unmanned aerial vehicles (UAVs) \cite{Gupta2016, zeng2017, zhang2019}, reconfigurable intelligent surfaces (RISs) \cite{macro42020, yan22020, yan32020, macro72020, macro52020} and so on. Due to UAVs' high mobility, they can be flexibly deployed to enhance the communication quality, while conventional terrestrial base stations (BSs) only serve the ground users in a fixed area. In addition, UAVs usually fly at a high altitude compared with the terrestrial infrastructure, which makes the transmission links between the UAV and the ground devices line-of-sight (LoS) dominated \cite{Schober2020}. Thanks to these advantages, UAVs are expected to play a key role in beyond fifth generation (B5G) and sixth generation (6G) networks \cite{zeng2019, zhu2020}. In the majority of research on UAV communications, secrecy is one of the key research aeras, in which authors focus on enhancing the secure communication quality via the joint optimization of the UAV trajectory and communication resource allocation. For instance, the authors of \cite{zhang2019} considered a simplified secure UAV communication system and maximized the average secrecy rate of the system via joint trajectory and power control design. The authors of \cite{lian2019} used a UAV as a jammer to transmit interfering signals to an eavesdropper, so as to improve the secrecy rate performance. A novel UAV-enabled secure communication system with cooperative jamming has been studied in \cite{zhong2019, lee2018, cai2018, zhou2019}, where one UAV acts as the legitimate transmitter and sends confidential data to the users, while another UAV acts as the jammer that delivers artificial noise (AN) to the eavesdroppers to weaken the quality of the eavesdropping channels. In particular, in \cite{lee2018, cai2018, zhou2019}, the scenario with multiple users and eavesdroppers was investigated, where the authors aimed at 
7
 maximizing the minimum secrecy rate among the legitimate users by jointly optimizing the UAV trajectory and corresponding communication resource allocation. In addition, the robust trajectory and transmit power design were studied in \cite{cui22018}, and the $\mathcal{S}$-Procedure method was used to efficiently solve the location uncertainty of the eavesdropper.



Benefiting from improving the propagation environment and enhancing the signal strength, RISs have been widely investigated as the technology enabler for realizing smart radio environments in the near future \cite{macro2020smart, macro2019smart, macro62020, yuan2020}. In general, an RIS is comprised of energy-efficient and cost-effective reconfigurable passive elements. Each element of the RIS can induce a phase shift on the incident signal by using a smart controller. Hence, with the aid of an RIS, the signals from different communication links can be added coherently at the desired receiver to enhance the received signal energy or can be added destructively at undesired receivers to avoid the information leakage \cite{wu2019intelligent}. This is also called passive beamforming. Due to the peculiar property of modifying the wireless propagation environment, RIS-assisted secure communication systems have attracted much attention \cite{guan2020,yu2019globe, xu2020, lu2020, yu2019robust}. In \cite{guan2020}, the authors investigated a simplified RIS-aided secure communication system, where the BS delivered confidential data to the user, while the eavesdropper intended to intercept the legitimate information. The RIS was utilized to enhance the quality of the legitimate links and weaken that of the wiretap links. By applying semidefinite relaxation (SDR) and Gaussian randomization methods, the authors maximized the achievable secrecy rate via jointly optimizing passive beamforming and transmit beamforming with AN. Since the SDR methods may not provide a rank-one solution, the majorization minimization (MM) technique \cite{yu2019globe} and manifold optimization theory \cite{xu2020} were used to obtain a rank-one solution. Robust and secure RIS-assisted communication systems have been studied in \cite{lu2020, yu2019robust}. In \cite{lu2020}, by capitalizing on the robust joint design of active beamforming and passive beamforming, the worst case of achievable secrecy rate was maximized for the colluding and non-colluding eavesdropping scenarios. The authors of \cite{yu2019robust} considered a secure wireless system comprised of multiple ground users, eavesdroppers, and RISs. It was assumed that the channel state information (CSI) of the eavesdropping channels was not perfectly known at the BS. Hence, a joint and robust design of the beamforming (including active beamforming at the BS and passive beamforming at the RISs) and the AN covariance matrix was proposed to maximize the system sum-rate under a given information leakage threshold.

From the above discussion, UAVs can provide LoS dominant transmission links with the ground users, thanks to their high mobility, while RISs can achieve passive beamforming by adjusting their reflecting elements smartly. Recently, the design of RIS-assisted UAV communication systems has attracted increasing attention \cite{macro32020, li2020, qian2019, long2020, wang2020joint, ge2020, hua2020}. In \cite{macro32020}, an RIS was utilized to assist the UAV relay system, and the simulation results demonstrated that deploying RISs could significantly improved the coverage and reliability of UAV communication systems. In \cite{li2020}, a UAV was used as the mobile BS to serve the ground user with the assistance of an RIS. The authors aimed at maximizing the average achievable rate by the joint optimization of the UAV trajectory and RIS's passive beamforming, and derived a closed-form solution of the RIS's phase-shift matrix for any given UAV trajectory. In \cite{qian2019}, an RIS was placed on the UAV to assist the users whose LoS path is blocked. Then, an efficient algorithm based on the reinforcement learning technique was proposed to solve the DL transmission capacity maximization problem. Similarly, in \cite{long2020}, a UAV equipped with an RIS was leveraged to achieve uplink secure communications. Based on the reinforcement learning method, the authors of \cite{wang2020joint} proposed a deep Q-network (DQN)-based algorithm to design the UAV's trajectory and RIS's passive beamforming to maximize the weighted fairness and data rate among all users. Additionally, the authors of \cite{ge2020} accounted for multiple RISs and a multi-antenna UAV, and maximized the received power by jointly optimizing passive beamforming, active beamforming, and the UAV’s trajectory. The authors of \cite{hua2020} studied a UAV-assisted RIS symbiotic radio system, where the UAV helped multiple RISs for their own information transmission. Based on statistical CSI, the problems of maximizing the minimum average rate and the weight sum rate over all RISs were solved, respectively, by the joint design of the UAV trajectory, RISs' passive beamforming, and RISs' scheduling. 

\begin{figure}[htp]
    \centering
    \includegraphics[width=3.0in]{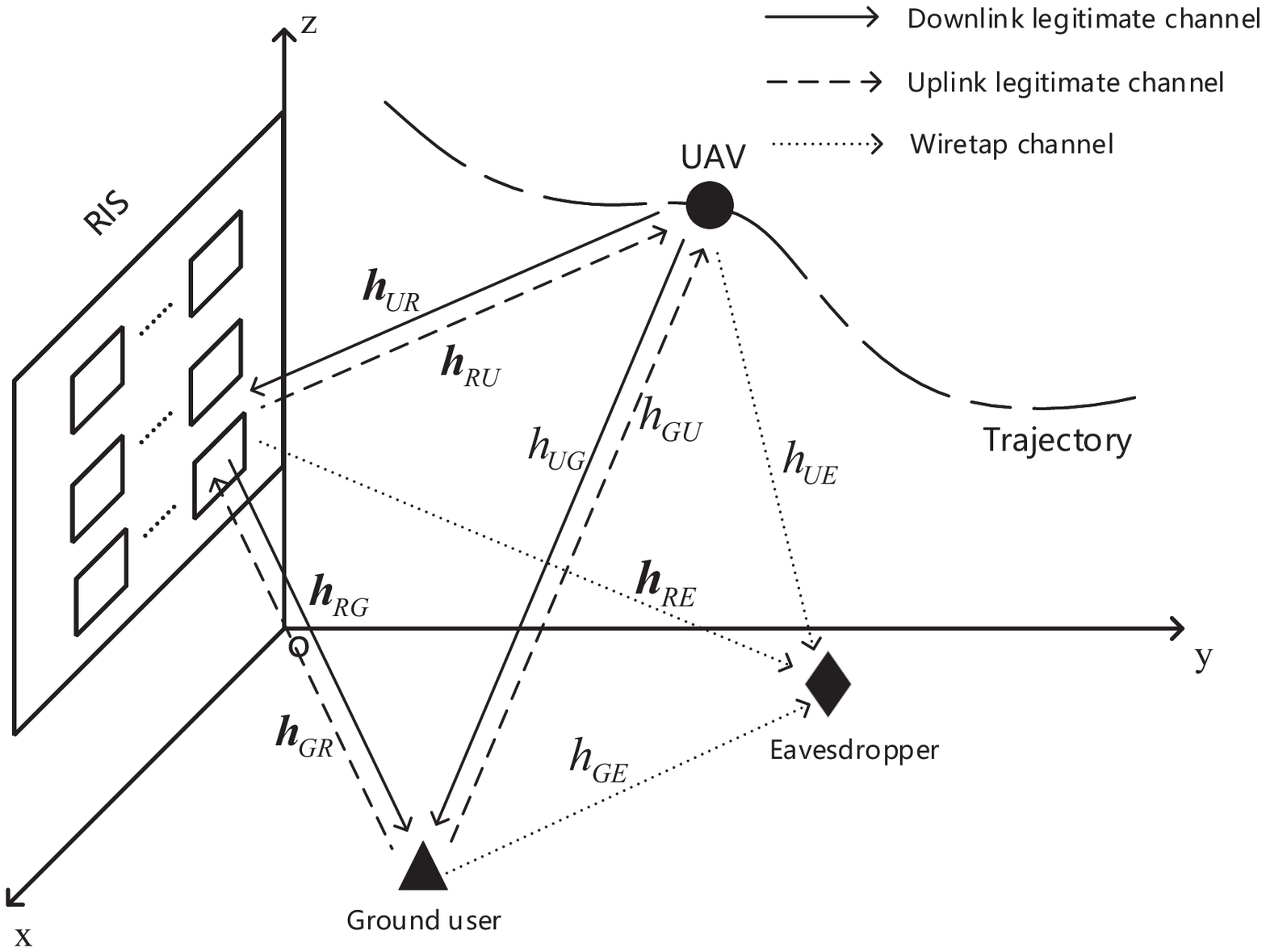}
    \caption{A RIS-assisted UAV secure communication system.}
    \label{figure_1}
\end{figure}

It is observed that among the current works on RIS-aided UAV communications, there exists very limited research on the design of secure communication systems. Furthermore, in the existing RIS-aided UAV secure communication systems, it is assumed that the perfect CSI of the eavesdropping channels is known. This assumption may not be easy to be met in practice since the eavesdroppers always avoid being detected by the legitimate transmitters, so as to intercept the legitimate information transmission successfully. Motivated by this, in this paper, we investigate a novel RIS-aided UAV secure communication system as shown in Fig.~\ref{figure_1}, where the UAV flies over a given flight period to serve the ground user, and the ground user also uploads some messages to the UAV, while a potential eavesdropper intends to wiretap their communications. However, in complex urban environment, the quality of the secure information transmission may be poor. Thus, an RIS is leveraged to enhance the communication quality of the legitimate links and weaken that of the eavesdropping links. Specifically, the entire flight duration is divided into time slots. We assume that the time division multiple access (TDMA) protocol is applied. As a result, we divide each time slot into two parts, i.e., one for the downlink (DL) transmission and the other for the uplink (UL) transmission\footnote{\textcolor{black}{In our paper, we assume that before the UAV starts moving, the ground user has achieved the uplink timing synchronization and has been allocated the corresponding radio resources by using random access (RA) schemes \cite{access1} generally used in the long term evolution (LTE) and LTE advanced (LTE-A).}}, where the UAV and the ground user are the legitimate transmitter (receiver) and receiver (transmitter), respectively. Since the eavesdropper always avoids to be detected as possible as it can, accurate estimates of the CSI of the eavesdropping links are usually not available. Hence, we assume imperfect CSI acquisition of the eavesdropping channels\footnote{\textcolor{black}{Although a passive eavesdropper always stays silent, it may send signals to its dedicated wireless system. In this case, the signal leakage from the eavesdropper to the legitimate transmitter can be used for channel estimation \cite{yu2019robust}. Furthermore, it is demonstrated that the eavesdropper’s location can be practically detected via a UAV-mounted camera or radar \cite{vision1,vision2}}.} and use a deterministic model \cite{schober2014} to describe the CSI uncertainty. Under these assumptions, a robust joint design of the UAV's trajectory, RIS's passive beamforming, and the transmit power control of the legitimate transmitter is formulated as a non-convex joint UL/DL optimization problem for maximizing the average worst-case secrecy rate. The considered problem is difficult to solve due to its non-convexity. We first address its non-smooth objective function and transform the formulated problem into an equivalent problem based on the results in \cite{zhang2019}. For the reformulated problem, however the corresponding optimization variables are coupled, which leads to a non-convex optimization problem. To tackle this difficulty, we propose an efficient algorithm based on the alternating optimization (AO) technique. More precisely, the reformulated problem is divided into three sub-problems: 1) transmit power control design for a given UAV trajectory and passive beamforming design; 2) passive beamforming design for a given UAV trajectory and transmit power control design; 3) UAV trajectory design for a given passive beamforming and transmit power control design. For sub-problem 1, we compute the optimal transmit power control design according to the special structure of the objective function. Then, for sub-problem 2, we utilize the $\mathcal{S}$-Procedure and successive convex approximation (SCA) techniques to handle the CSI uncertainty and the non-concave objective function, respectively. Finally, it is challenging to account for the CSI uncertainty and the small-scale fading component of the channel between the UAV and RIS. To cope with these challenges, we use the UAV trajectory of the previous iteration to estimate the current small-scale fading component of the channel between the UAV and RIS and the worst-case setup of the eavesdropping links. Then, the SCA method is applied to solve the optimization sub-problem efficiently. Simulation results demonstrate that our proposed algorithm can significantly increase the average secrecy rate, as compared to benchmark algorithms.

The remainder of this paper is organized as follows. In Section II, we present the system model and problem formulation. In Section III, we propose efficient algorithms based on the AO technique to solve the formulated joint UL/DL optimization problem. Simulation results are illustrated in Section IV. Finally, we conclude the paper in Section V.

$\mathcal{}$

\section{System Model and Problem Formulation}
\subsection{System Model}
    As shown in Fig.~\ref{figure_1}, we consider a UAV-enabled communication system where a rotary-wing UAV and a ground user communicate with each other, while an eavesdropper attempts to intercept their legitimate communications. Due to the limited capacity of the UAV, the performance of such secure communication may be low. Thus, we use a building-mounted RIS to assist the secure data transmission. \textcolor{black}{It is assumed that control and non-payload communications (CNPC) links\footnote{\textcolor{black}{CNPC links \cite{control2016} with stringent latency and security requirements are established in the UAV system for supporting safety-critical functions, such as real-time control, collision and crash avoidance, and so on. From \cite{Mun2020}, it is known that the CNPC technique has been widely recognized, and the corresponding development and standardization have been launched.}} are constructed between the UAV, the RIS and the ground user for the transmission of the control signals.} Without loss of generality, we assume that all communication nodes are placed in the three dimensional (3D) Cartesian coordinate system. The ground user's and the eavesdropper's horizontal coordinates are denoted by $\mathbf{w}_G=[x_G, y_G]^T$ and $\mathbf{w}_E=[x_E, y_E]^T$, respectively. The UAV is assumed to fly at a constant altitude denoted by $z_U$ for a given flight period denoted by \emph{T}. For tractability, \emph{T} is discretized into \emph{N} time slots, namely, $T = N\delta_t$, where $\delta_t$ is the time slot length. Therefore, the UAV's time-varying horizontal trajectory is denoted as the sequence $\mathbf{q}[n]=[x[n], y[n]]^T, n \in \mathcal{N}\triangleq{\{1,\cdots, N \}}$, which should meet the following mobility constraints:
    \begin{subequations} \label{mobility.1}
    \begin{align}
        & ||\mathbf{q}[n+1]-\mathbf{q}[n]||^{2} \leq D^2,n=1,\cdots,N-1, \label{mobility.1.a}\\
        & ||\mathbf{q}[N]-\mathbf{q}_F||^{2} \leq D^2, \mathbf{q}[1] = \mathbf{q}_0, \label{mobility.1.b}
    \end{align}
    \end{subequations}
    where $\mathbf{q}_0$ and $\mathbf{q}_F$ are the predetermined initial and final horizontal locations of the UAV, respectively, $D=v_{\max}\delta_t$ is the maximum horizontal distance that the UAV can fly in $\delta_t$, and $v_{\max}$ is the maximum speed of the UAV. We assume that the UAV, the ground user, and the eavesdropper are equipped with a single-antenna. The RIS is equipped with $M=M_x\times M_z$ reflecting elements, forming an $M_x\times M_z$ uniform rectangular array (URA), and a controller that can intelligently adjust the phase shift of each element. The RIS is located in the $x$-$z$ plane, and its altitude and horizontal coordinates are denoted by $z_R$ and $\mathbf{w}_R = [x_R, y_R]^T$, respectively. \textcolor{black}{${\rm diag}\left(\boldsymbol{x}\right)$ denotes a diagonal matrix in which each diagonal element is the corresponding element in $\boldsymbol{x}$.} The diagonal phase-shift matrix for the RIS in the time slot $n$ is denoted by $\boldsymbol{\Theta}[n]={\rm diag}\left( e^{j\theta_1[n]}, e^{j\theta_2[n]},\cdots, e^{j\theta_M[n]}\right)$, where $\theta_i[n] \in \left[0,2\pi\right), i \in \mathcal{M}\triangleq{\{1,\cdots, M\}}$, is the phase shift of the $i$th reflecting element within a single time slot.
    
    To ensure the mutual communication between the UAV and the ground user, we assume that the TDMA protocol is applied for them. Specifically, we utilize a weighted factor $w \in \left[0, 1\right]$ to divide a single flight time slot into two parts: $w\delta_t$ for the DL transmission in which the UAV serves the ground user, and $\left(1 - w\right)\delta_t$ for the UL transmission in which the ground user uploads data that the UAV intends to harvest. The details are specified as follows.
    
    \subsubsection{DL Transmission} In this case, the UAV and the ground node serve as the legitimate transmitter and receiver, respectively. Let $p[n]$ denote the transmit power of the UAV in time slot $n$. In practice, $p[n]$ is usually subject to both average and peak limits over time, denoted by $\Bar{P}$ and $P_{peak}$, respectively. Thus, the transmit power constraints are expressed as
    \begin{subequations}\label{power.1}
        \begin{align}
        &\frac{1}{N}\sum\limits_{n=1}^N p[n] \leq \Bar{P},\label{power.1.a}\\
        &0 \leq p[n] \leq P_{peak}, \forall n.\label{power.1.b}
    \end{align}
    \end{subequations}
   
    We assume the Rician fading channel model for all communication links. Hence, the small-scale fading component of the link from the UAV to the RIS (U-R link) in the $n$th time slot, denoted by $\boldsymbol{h}_{U\!R}[n] \in \mathbb{C}^{M\times 1}$, can be expressed as
    \begin{equation}
        \boldsymbol{h}_{U\!R}[n] = \sqrt{\frac{\beta_{U\!R}}{1+\beta_{U\!R}}}\boldsymbol{h}_{U\!R}^{\rm{LoS}}[n] + \sqrt{\frac{1}{1+\beta_{U\!R}}}\boldsymbol{h}_{U\!R}^{\rm{NLoS}},
    \end{equation}
    where $\beta_{U\!R}$ is the Rician factor of the U-R link, $\boldsymbol{h}_{U\!R}^{\rm{LoS}}[n]$ is the deterministic LoS component, and $\boldsymbol{h}_{U\!R}^{\rm{NLoS}}$ is the non-LoS (NLoS) component which is modeled by the circularly symmetric complex Gaussian (CSCG) distribution with zero mean and unit variance. In particular, $\boldsymbol{h}_{U\!R}^{\rm{LoS}}[n]$ depends on the UAV trajectory at time slot $n$, and it can be expressed as \cite{Tse2009Fundamentals, zhu2019}
    \begin{equation}\label{LoScomponent.1}
        \boldsymbol{h}_{U\!R}^{\rm{LoS}}[n]=\boldsymbol{a}_y[n]\otimes \boldsymbol{a}_x[n],
    \end{equation}
        where 
        \begin{align}
            \boldsymbol{a}_x[n]&=\left[1, e^{-j\!\frac{2\pi}{\lambda}d\cos{\phi_{U\!R}[n]}\sin{\varphi_{U\!R}[n]}}, ... , \right.\nonumber\\
            &\quad\quad\quad \left. e^{ -j\!\frac{2\pi}{\lambda}\!(M_x-1)d\cos{\phi_{U\!R}[n]}\sin{\varphi_{U\!R}[n]}} \right]^T,\nonumber\\
            \boldsymbol{a}_y[n]&=\left[1, e^{-j\!\frac{2\pi}{\lambda}d\sin{\phi_{U\!R}[n]}\sin{\varphi_{U\!R}[n]}}, ... , \right.\nonumber\\
            &\quad\quad\quad \left. e^{ -j\!\frac{2\pi}{\lambda}\!(M_z-1)d\sin{\phi_{U\!R}[n]}\sin{\varphi_{U\!R}[n]}} \right]^T,\nonumber\\
            &~\sin{\phi_{U\!R}[n]}\sin{\varphi_{U\!R}[n]}=\frac{z_U-z_R}{d_{U\!R}[n]},\nonumber\\
            &\cos{\phi_{U\!R}[n]}\sin{\varphi_{U\!R}[n]}=\frac{x_R-x[n]}{d_{U\!R}[n]},\nonumber
        \end{align}
     $d_{U\!R}[n]=\sqrt{(z_U-z_R)^2+||\mathbf{q}[n]-\mathbf{w}_R||^2}$ denotes the distance between the UAV and the RIS in the $n$th time slot, $\phi_{U\!R}[n]$ and $\varphi_{U\!R}[n]$ represent the azimuth and elevation angles of the LoS component in time slot $n$, respectively, $d$ is the antenna separation, and $\lambda$ is the carrier wavelength. The small-scale fading components of the links from the RIS to the ground user (R-G link), the RIS to the eavesdropper (R-E link), the UAV to the ground user (U-G link), and the UAV to the eavesdropper (U-E link) can be generated with a similar procedure, and they are denoted as $\boldsymbol{h}_{RG}^H \in \mathbb{C}^{1\times M}$,  $\boldsymbol{h}_{RE}^H \in \mathbb{C}^{1\times M}$, $h_{U\!G} \in \mathbb{C}$, and $h_{U\!E} \in \mathbb{C}$, respectively. We use the distance-dependent path loss model in \cite{macro22020}, \cite{yan2020} for the reflected links, i.e., the links from the UAV to the ground user via the RIS (U-R-G link) and the UAV to the eavesdropper via the RIS (U-R-E link), which can be expressed as
    \begin{align}\label{pathloss.1}
        L_{U\!RG}[n]&=\sqrt{\rho \left(d_{U\!R}[n]d_{RG}\right)^{-\alpha}},\\ L_{U\!RE}[n]&=\sqrt{\rho \left(d_{U\!R}[n]d_{RE}\right)^{-\alpha}},
    \end{align}
    where 
    \begin{align}
        d_{RG} = \sqrt{z_R^2+||\mathbf{w}_R-\mathbf{w}_G||^2},\nonumber\\
        d_{RE} = \sqrt{z_R^2+||\mathbf{w}_R-\mathbf{w}_E||^2},\nonumber
    \end{align}
    $\rho$ is the path loss at the reference distance $D_0=1$ m, and $\alpha$ is the path loss exponent for the U-R-G and the U-R-E links. For the direct links, namely, the U-G and the U-E links, the corresponding distance-dependent path loss models are given by
    \begin{align}\label{pathloss.2}
        L_{U\!G}[n]=\sqrt{\rho \left(z_U^2+||\mathbf{q}[n]-\mathbf{w}_G||^2\right)^{-\frac{\kappa}{2}}},\\
         L_{U\!E}[n]=\sqrt{\rho \left(z_U^2+||\mathbf{q}[n]-\mathbf{w}_E||^2\right)^{-\frac{\kappa}{2}}},
    \end{align}
    where $\kappa$ is the path loss exponent for the U-G and the U-E links.
    
    With the above channel models, the received signal-to-noise ratios (SNRs) of the ground user and the eavesdropper in the $n$th time slot can be respectively expressed as
    \begin{align}
        \gamma_{U\!G}[n]&=\frac{p[n]{\left|L_{UG}[n]h_{U\!G} + L_{U\!RG}[n]\boldsymbol{h}_{RG}^H \boldsymbol{\Theta}_{d}[n] \boldsymbol{h}_{U\!R}[n]\right|^2}}{\sigma^2},\\
        \gamma_{U\!E}[n]&=\frac{p[n]{\left|L_{U\!E}[n]h_{U\!E} + L_{U\!RE}[n]\boldsymbol{h}_{RE}^H \boldsymbol{\Theta}_{d}[n] \boldsymbol{h}_{U\!R}[n]\right|^2}}{\sigma^2},
    \end{align}
    where $\boldsymbol{\Theta}_{d}[n]={\rm diag}\left( e^{j\theta^d_1[n]}, e^{j\theta^d_2[n]},\cdots, e^{j\theta^d_M[n]}\right)$ is the phase-shift matrix of the DL transmission in the time slot $n$ and $\sigma^2$ is the noise variance. Thus, the achievable rates in bits/second/Hertz (bps/Hz) at the ground user and the eavesdropper in time slot $n$ are respectively given by
    \begin{subequations}\label{rate.1}
    \begin{align}
        R_{U\!G}[n]&=\log_2(1+\gamma_{U\!G}[n]),\label{rate.1.a}\\
        R_{U\!E}[n]&=\log_2(1+\gamma_{U\!E}[n]).\label{rate.1.b}
    \end{align}
    \end{subequations}
    \subsubsection{UL Transmission} In this case, the ground user and the UAV serve as the legitimate transmitter and receiver, respectively. Denote by $g[n]$ the transmit power of the ground user in time slot $n$. Similarly, $g[n]$ is constrained by an average power limit $\Bar{G}$ and a peak power limit $G_{peak}$, i.e.,
    \begin{subequations}\label{power_g.1}
    \begin{align}
        &\frac{1}{N}\sum\limits_{n=1}^N g[n] \leq \Bar{G},\label{power_g.1.a}\\
        &0 \leq g[n] \leq G_{peak}, \forall n.\label{power_g.1.b}
    \end{align}
    \end{subequations}
    Since the ground user and the eavesdropper are both on the ground, we assume that the eavesdropping channel between the ground user and the eavesdropper (G-E link) is modeled as a Rayleigh fading channel. Thus, the small-scale fading component of the G-E link, denoted by $h_{GE}$, is assumed to be a zero-mean and unit-variance CSCG random variable. The distance-dependent path loss of the G-E link is given by
    \begin{equation}
        L_{G\!E} = \sqrt{\rho \left(||\mathbf{w}_G-\mathbf{w}_E||^2\right)^{-\frac{\varsigma}{2}}},
    \end{equation}
    where $\varsigma$ is the path loss exponent related to the G-E link. Similar to the DL transmission, the other channels in the UL transmission are assumed to be Rician distributed, and thus, we omit their specific structures for brevity. The small-scale fading components of the links from the RIS to the UAV, the ground user to the RIS, and the ground user to the UAV are denoted as $\boldsymbol{h}_{RU}^H[n] \in \mathbb{C}^{1\times M}$, $\boldsymbol{h}_{G\!R} \in \mathbb{C}^{M\times 1}$, $h_{GU} \in \mathbb{C}$, respectively. We still use $L_{U\!G}[n]$ and $L_{U\!RG}[n]$ to express the distance-dependent path loss models of the user-RIS-UAV (G-R-U) link and user-UAV (G-U) link, respectively. Therefore, the received SNRs of the UAV and the eavesdropper in the $n$th time slot can be respectively written as
    \begin{align}
        \gamma_{GU}[n]&=\frac{g[n]{\left|L_{U\!G}[n]h_{GU} + L_{U\!RG}[n]\boldsymbol{h}_{RU}^H[n] \boldsymbol{\Theta}_{u}[n]\boldsymbol{h}_{G\!R} \right|^2}}{\sigma^2},\\
        \gamma_{G\!E}[n]&=\frac{g[n]{\left|L_{G\!E}h_{GE} + L_{G\!R\!E}\boldsymbol{h}_{RE}^H \boldsymbol{\Theta}_{u}[n] \boldsymbol{h}_{G\!R}\right|^2}}{\sigma^2},
    \end{align}
    where 
    $$L_{G\!R\!E} = \sqrt{\rho \left[\left(z_R^2+||\mathbf{w}_R-\mathbf{w}_G||^2\right)\left(z_R^2+||\mathbf{w}_R-\mathbf{w}_E||^2\right)\right]^{-\frac{\alpha}{2}}}$$ is the large-scale fading component of the user-RIS-eavesdropper (G-R-E) link, and $\boldsymbol{\Theta}_{u}[n]={\rm diag}\left( e^{j\theta^u_1[n]}, e^{j\theta^u_2[n]},\cdots, e^{j\theta^u_M[n]}\right)$ is the phase-shift matrix of the uplink transmission in the time slot $n$. Hence, the achievable rates in bps/Hz from the ground user to the UAV and the eavesdropper in time slot $n$ are respectively given by
    \begin{subequations}\label{rate.2}
    \begin{align}
        R_{GU}[n]&=\log_2(1+\gamma_{GU}[n]),\label{rate.2.a}\\
        R_{G\!E}[n]&=\log_2(1+\gamma_{G\!E}[n]).\label{rate.2.b}
    \end{align}
    \end{subequations}
    
\subsection{CSI Assumption}
In general, the legitimate transmitter is able to periodically update and refine the CSI of the legitimate receiver based on uplink pilots. In addition, some channel estimation techniques \cite{he2020, Mishra2019, liuhang2020} have been proposed for CSI acquisition in the presence of RISs recently. Based on these considerations, we assume that the CSI of the legitimate links is perfectly available in a central controller. However, the eavesdropper usually avoids being detected and tracked by the legitimate transmitter in order to intercept the legitimate communications. Hence, the estimated CSI of the eavesdropping channels are usually not accurate at the central controller. For this reason, we first rewrite $\gamma_{U\!E}[n]$ and $\gamma_{G\!E}[n]$ as 
\begin{align}
    \gamma_{U\!E}[n] &= \frac{p[n]}{\sigma^2}\left|\boldsymbol{h}_{E1}^H\mathbf{H}_{E1}[n]\boldsymbol{v}^{d}[n]\right|^2,\\
    \gamma_{G\!E}[n] &= \frac{g[n]}{\sigma^2}\left|\boldsymbol{h}_{E2}^H\mathbf{H}_{E2}\boldsymbol{v}^{u}[n]\right|^2,
\end{align}
where 
$$\mathbf{H}_{E1}[n] = {\rm diag} \left( \left[\begin{matrix}L_{U\!RE}[n]\boldsymbol{h}_{U\!R}[n] \\ L_{U\!E}[n] \end{matrix}\right]\right),$$
$$\mathbf{H}_{E2} = {\rm diag} \left( \left[\begin{matrix}L_{G\!RE}\boldsymbol{h}_{G\!R} \\ L_{G\!E} \end{matrix}\right]\right),$$
$\boldsymbol{h}_{E1}=\left[\boldsymbol{h}_{R\!E}^H, h_{U\!E}\right]^H$, $\boldsymbol{h}_{E2}=\left[\boldsymbol{h}_{R\!E}^H, h_{G\!E}\right]^H$, and $\boldsymbol{v}^{d}[n]=\left[v^{d}_1[n], v^{d}_2[n], \cdots, v^{d}_M[n], 1\right]^T$ ($v^{d}_i[n]=e^{j\theta^{d}_i[n]}, \forall n, i$). The structure of $\boldsymbol{v}^{u}[n]$ is similar to $\boldsymbol{v}^{d}[n]$. In particular, the links related to the eavesdropper are $\boldsymbol{h}_{E1}$ and $\boldsymbol{h}_{E2}$. Then, we utilize a deterministic model to characterize the CSI uncertainty. \textcolor{black}{Let $\lVert \boldsymbol{x} \rVert$ denote the Euclidean norm of the complex-valued vector $\boldsymbol{x}$}. The uncertainties of the eavesdropping channels in the DL and UL transmissions are respectively modeled as\footnote{\textcolor{black}{Since the UAV's location is time-varying, the worst-case setup of the eavesdropping channels is also time-varying along with the UAV's location. Thus, we add time slot index $n$ to change $\Delta\boldsymbol{h}_{E1}$ to $\Delta\boldsymbol{h}_{E1}[n]$. Similarly, we change $\Delta\boldsymbol{h}_{E2}$ to $\Delta\boldsymbol{h}_{E2}[n]$, $\boldsymbol{h}_{E1}$ to $\boldsymbol{h}_{E1}[n]$, and $\boldsymbol{h}_{E2}$ to $\boldsymbol{h}_{E2}[n]$.}}
\textcolor{black}{
\begin{subequations}\label{uncertainty.1}
\begin{align}
    &\boldsymbol{h}_{E1}[n] = \Bar{\boldsymbol{h}}_{E1} + \Delta \boldsymbol{h}_{E1}[n],\nonumber\\
    &\Omega_1\triangleq{\left\{\Delta \boldsymbol{h}_{E1}[n] \in \mathbb{C}^{M+1 \times 1}: \lVert\Delta \boldsymbol{h}_{E1}[n] \rVert \leq \epsilon_1, \forall n\right\}}, \label{uncertainty.1.a}\\
    &\boldsymbol{h}_{E2}[n] = \Bar{\boldsymbol{h}}_{E2} + \Delta \boldsymbol{h}_{E2}[n],\nonumber\\
    &\Omega_2\triangleq{\left\{\Delta \boldsymbol{h}_{E2}[n] \in \mathbb{C}^{M+1 \times 1}: \lVert\Delta \boldsymbol{h}_{E2}[n] \rVert \leq \epsilon_2, \forall n\right\}},\label{uncertainty.1.b}
\end{align}
\end{subequations}}where $\Bar{\boldsymbol{h}}_{E1}=\left[\Bar{\boldsymbol{h}}_{R\!E}^H, \Bar{h}_{U\!E}\right]^H$ and $\Bar{\boldsymbol{h}}_{E2}=\left[\Bar{\boldsymbol{h}}_{R\!E}^H, \Bar{h}_{G\!E}\right]^H$ are the estimated CSI, and $\Delta \boldsymbol{h}_{E1}[n]$ and $\Delta \boldsymbol{h}_{E2}[n]$ represent the estimated errors for $\Bar{\boldsymbol{h}}_{E1}$ and $\Bar{\boldsymbol{h}}_{E2}$, respectively. The continuous sets $\Omega_1$ and $\Omega_2$ contain all possible CSI uncertainties with norms bounded by the uncertainty radii $\epsilon_1$ and $\epsilon_2$, respectively.

\subsection{Problem Formulation}
Based on \eqref{rate.1} and \eqref{rate.2}, the worst-case secrecy rates in time slot $n$ in the DL and UL transmissions can be respectively expressed as
\begin{subequations}\label{rate.3}
\begin{align}
    R_{sec}^{down}[n]&=\left[R_{U\!G}[n]-\max_{\Delta \boldsymbol{h}_{E1}[n] \in \Omega_1}R_{U\!E}[n]\right]^+,\label{rate.3.a}\\
    R_{sec}^{up}[n]&=\left[R_{GU}[n]-\max_{\Delta \boldsymbol{h}_{E2}[n] \in \Omega_2}R_{G\!E}[n]\right]^+,\label{rate.3.b}
\end{align}
\end{subequations}
where $[x]^+ \triangleq{\max(x, 0)}$. Hence, the average worst-case secrecy rate of the joint UL/DL RIS-assisted UAV secure communication system is given by
\begin{equation}\label{rate.4}
    R_{sec} = \frac{1}{N}\sum\limits_{n=1}^N \left\{w R_{sec}^{down}[n] + \left(1-w\right) R_{sec}^{up}[n]\right\}.
\end{equation}
Our objective is to maximize $R_{sec}$ by jointly optimizing the UAV's trajectory $\mathbf{Q}\triangleq \{\mathbf{q}[n], n\in\mathcal{N}\}$, the phase-shift matrices $\boldsymbol{\Phi}_{d}\triangleq \{\boldsymbol{\Theta}_{d}[n], n\in\mathcal{N}\}$ and $\boldsymbol{\Phi}_{u}\triangleq \{\boldsymbol{\Theta}_{u}[n], n\in\mathcal{N}\}$ of the RIS, the UAV's transmit power $\mathbf{p}\triangleq \{p[n], n\in\mathcal{N}\}$, and the transmit power $\mathbf{g}\triangleq \{g[n], n\in\mathcal{N}\}$ of the ground user. Therefore, the problem can be formulated as
\begin{subequations}\label{optimal.1}
\begin{align}
     &\max\limits_{\mathbf{Q}, \boldsymbol{\Phi}_d, \boldsymbol{\Phi}_u
     \atop
     \mathbf{p}, \mathbf{g}} R_{sec}\label{optimal.1.a}\\
     & ~~~\textrm{s.t.} \quad~ 0 \leq \theta^d_i[n] < 2\pi, \forall n,i,\label{optimal.1.b} \\
     & \quad \quad\quad~~ 0 \leq \theta^u_i[n] < 2\pi, \forall n,i,\label{optimal.1.c} \\
     & \quad \quad\quad~~ \eqref{mobility.1}, \eqref{power.1}, \eqref{power_g.1}. \nonumber
 \end{align}
\end{subequations}
It is observed that the constraints of problem \eqref{optimal.1} are all convex. However, it is still difficult to solve problem \eqref{optimal.1} since the objective function of problem \eqref{optimal.1} is highly non-concave with respect to $\mathbf{Q}$, $\boldsymbol{\Phi}_{d}$, $\boldsymbol{\Phi}_{u}$, $\mathbf{p}$, and $\mathbf{g}$. In the next section, we develop an efficient algorithm to solve problem \eqref{optimal.1}.


\section{Proposed Solution for Joint UL/DL Optimization}
In this section, we focus on solving the joint UL/DL optimization problem \eqref{optimal.1}. Based on Lemma 1 in \cite{zhang2019}, it is known that the transmit power control design can guarantee $R_{U\!G}[n]-\max \limits_{\Delta \boldsymbol{h}_{E1} \in \Omega_1}R_{U\!E}[n] \ge 0$ and $R_{GU}[n]-\max \limits_{\Delta \boldsymbol{h}_{E2} \in \Omega_2}R_{G\!E}[n] \ge 0$, since the optimal transmit power of the UAV and the ground user in time slot $n$, denoted as $p^{op}[n]$ and $g^{op}[n]$, respectively, are zero once
the quality of the eavesdropping channels is better than that of the legitimate channels in time slot $n$. Therefore, we reformulate problem \eqref{optimal.1} as
\begin{align}\label{optimal_r.1}
     & \max\limits_{\mathbf{Q}, \boldsymbol{\Phi}_d, \boldsymbol{\Phi}_u
     \atop
     \mathbf{p}, \mathbf{g}} ~~\frac{1}{N}\sum\limits_{n=1}^N \left\{w \Tilde{R}_{sec}^{down}[n] + \left(1-w\right) \Tilde{R}_{sec}^{up}[n]\right\}\\
     & ~~~\textrm{s.t.} \quad~~ \eqref{mobility.1}, \eqref{power.1}, \eqref{power_g.1}, \eqref{optimal.1.b}, \eqref{optimal.1.c}, \nonumber
 \end{align}
where 
$$\Tilde{R}_{sec}^{down}[n]=\left[R_{U\!G}[n]-\max_{\Delta \boldsymbol{h}_{E1}[n] \in \Omega_1}R_{U\!E}[n]\right]$$ and $$\Tilde{R}_{sec}^{up}[n]=\left[R_{GU}[n]-\max_{\Delta \boldsymbol{h}_{E2}[n] \in \Omega_2}R_{G\!E}[n]\right].$$As a result, the non-smoothness of problem \eqref{optimal.1} is addressed, and there exists no performance loss in this step. However, problem \eqref{optimal_r.1} is still difficult to solve due to the coupled optimization variables $\mathbf{Q}$, $\boldsymbol{\Phi}_{d}$, $\boldsymbol{\Phi}_{u}$, $\mathbf{p}$, and $\mathbf{g}$ in the objective function. To cope with this difficulty, we propose an efficient algorithm based on the AO method. Specifically, we divide problem \eqref{optimal_r.1} into three sub-problems: 
\begin{itemize}
    \item [1)]
    The optimization of the transmit power $\mathbf{p}$ and $\mathbf{g}$ under the given UAV trajectory $\mathbf{Q}$ and phase-shift matrices $\boldsymbol{\Phi}_d$ and $\boldsymbol{\Phi}_u$ (referred to as sub-problem 1);
    \item [2)]
    The optimization of the phase-shift matrices $\boldsymbol{\Phi}_d$ and $\boldsymbol{\Phi}_u$ under the given UAV trajectory $\mathbf{Q}$ and transmit power $\mathbf{p}$ and $\mathbf{g}$ (referred to as sub-problem 2);
    \item [3)]
    The optimization of the UAV trajectory $\mathbf{Q}$ under the given phase-shift matrices $\boldsymbol{\Phi}_d$ and $\boldsymbol{\Phi}_u$ and transmit power $\mathbf{p}$ and $\mathbf{g}$ (referred to as sub-problem 3).
\end{itemize}
The details are presented in the next three subsections, and subsequently the overall algorithm is summarized.

\subsection{Solution to Sub-Problem 1}
For any given $\mathbf{Q}$, $\boldsymbol{\Phi}_d$, and $\boldsymbol{\Phi}_u$, we have 
\begin{align}
    &\boldsymbol{h}_{G1}^H\mathbf{H}_{G1}[n]\boldsymbol{v}^d[n]=L_{U\!G}[n]h_{U\!G} + L_{U\!RG}[n]\boldsymbol{h}_{RG}^H \boldsymbol{\Theta}_d[n] \boldsymbol{h}_{U\!R}[n],\nonumber\\
    &\boldsymbol{h}_{G2}^H[n]\mathbf{H}_{G2}[n]\boldsymbol{v}^u[n]=L_{U\!G}[n]h_{GU} \nonumber\\
    &\quad\quad\quad\quad\quad\quad\quad\quad\quad~+ L_{U\!RG}[n]\boldsymbol{h}_{RU}^H[n] \boldsymbol{\Theta}_{u}[n]\boldsymbol{h}_{G\!R},\nonumber
\end{align}
where 
$$\mathbf{H}_{G1}[n] = {\rm diag}\left(\left[\begin{matrix}L_{U\!RG}[n]\boldsymbol{h}_{U\!R}[n] \\ L_{U\!G}[n] \end{matrix}\right]\right),$$
$$\mathbf{H}_{G2}[n] = {\rm diag}\left(\left[\begin{matrix}L_{U\!RG}[n]\boldsymbol{h}_{GR} \\ L_{U\!G}[n] \end{matrix}\right]\right),$$
$\boldsymbol{h}_{G1} = \left[\boldsymbol{h}_{RG}^H, h_{U\!G} \right]^H$, and $\boldsymbol{h}_{G2}[n] = \left[\boldsymbol{h}_{RU}^H[n], h_{GU} \right]^H$. Then, sub-problem 1 can be expressed as
\begin{align}\label{optpower.1}
    & \max\limits_{\mathbf{p}, \mathbf{g}} \quad \frac{1}{N}\sum\limits_{n=1}^N\left[ wR_{down}^{power}[n] +  \left(1-w\right) R_{up}^{power}[n]\right]\\
    & ~~\textrm{s.t.} ~~~~ \eqref{power.1}, \eqref{power_g.1},\nonumber
\end{align}where
\begin{equation}
    \begin{split}
        &R_{down}^{power}[n]=\log_2\left(1 + \frac{p[n]}{\sigma^2} \left|\boldsymbol{h}_{G1}^H\mathbf{H}_{G1}[n]\boldsymbol{v}^d[n]\right|^2\right)\\
    &\quad- \log_2\left(1+\max\limits_{\Delta\boldsymbol{h}_{E1}[n]\in\Omega_1}\frac{p[n]}{\sigma^2}\left|\boldsymbol{h}_{E1}^H[n]\mathbf{H}_{E1}[n]\boldsymbol{v}^d[n]\right|^2\right)\nonumber
    \end{split}
\end{equation}
and
\begin{equation}
    \begin{split}
        &R_{up}^{power}[n]=\log_2\left(1 + \frac{g[n]}{\sigma^2} \left|\boldsymbol{h}_{G2}^H[n]\mathbf{H}_{G2}[n]\boldsymbol{v}^u[n]\right|^2\right)\\
    &\quad -\log_2\left(1+\max\limits_{\Delta\boldsymbol{h}_{E2}[n]\in\Omega_2}\frac{g[n]}{\sigma^2}\left|\boldsymbol{h}_{E2}^H[n]\mathbf{H}_{E2}\boldsymbol{v}^u[n]\right|^2\right).\nonumber
    \end{split}
\end{equation}
Infinitely many possible CSI uncertainties in $\Omega_1$ and $\Omega_2$ make problem \eqref{optpower.1} intractable. However, the special structure of $\left|\boldsymbol{h}_{E1}^H[n]\mathbf{H}_{E1}[n]\boldsymbol{v}^d[n]\right|$ and $\left|\boldsymbol{h}_{E2}^H[n]\mathbf{H}_{E2}\boldsymbol{v}^u[n]\right|$ can be utilized to address this problem. Let $\arg\left(\boldsymbol{x}\right)$ denote the phase angle vector of $\boldsymbol{x}$, in which each element is the phase angle of the corresponding element in $\boldsymbol{x}$. We first have the following inequality:
\begin{align}
    \left|\boldsymbol{h}_{E1}^H[n]\mathbf{H}_{E1}[n]\boldsymbol{v}^d[n]\right| &\leq \left|\Bar{\boldsymbol{h}}_{E1}^H\mathbf{H}_{E1}[n]\boldsymbol{v}^d[n]\right| \nonumber\\
    &\quad+ \left|\Delta\boldsymbol{h}_{E1}^H[n]\mathbf{H}_{E1}[n]\boldsymbol{v}^d[n]\right|,\nonumber
\end{align}
where the equality holds if and only if $$\arg\left(\Bar{\boldsymbol{h}}_{E1}^H\mathbf{H}_{E1}[n]\boldsymbol{v}^d[n]\right) = \arg\left(\Delta\boldsymbol{h}_{E1}^H[n]\mathbf{H}_{E1}[n]\boldsymbol{v}^d[n]\right).$$ Thus, $\max\limits_{\Delta\boldsymbol{h}_{E1}[n]\in\Omega_1}p[n]\left|\boldsymbol{h}_{E1}^H[n]\mathbf{H}_{E1}[n]\boldsymbol{v}^d[n]\right|^2/\sigma^2$ can be transformed into
\begin{subequations}\label{opterror.1}
\begin{align}
    &\max\limits_{\Delta\boldsymbol{h}_{E1}[n]} ~ \left|\Delta\boldsymbol{h}_{E1}^H[n]\mathbf{H}_{E1}[n]\boldsymbol{v}^d[n]\right|^2\label{opterror.1.a}\\
    &~~\textrm{s.t.} ~~~ \lVert\Delta \boldsymbol{h}_{E1}[n] \rVert \leq \epsilon_1,\label{opterror.1.b}\\
    &\quad\quad~~ \arg\left(\Bar{\boldsymbol{h}}_{E1}^H\mathbf{H}_{E1}[n]\boldsymbol{v}^d[n]\right) = \arg\left(\Delta\boldsymbol{h}_{E1}^H[n]\mathbf{H}_{E1}[n]\boldsymbol{v}^d[n]\right).\label{opterror.1.c}
\end{align}
\end{subequations}
To facilitate the subsequent derivations, $\Delta\boldsymbol{h}_{E1}[n]$ can be expressed as
\begin{align}
    \Delta\boldsymbol{h}_{E1}[n] &= \left[\left|\Delta h_{E1,1}[n]\right|e^{j\tau_1[n]}, \left|\Delta h_{E1,2}[n]\right|e^{j\tau_2[n]}, \cdots, \right.\nonumber\\
    &\quad~\left.\left|\Delta h_{E1,M+1}[n]\right|e^{j\tau_{M+1}[n]}\right],
\end{align}
where $\left|\Delta h_{E1,k}[n]\right|$ and $\tau_k[n]$ are the magnitude and phase angle of the $k$th element of $\Delta\boldsymbol{h}_{E1}[n]$ in time slot $n$, respectively, and $k \in \mathcal{K} = \{1, \cdots, M+1\}$. Let $\boldsymbol{c}[n] = \mathbf{H}_{E1}[n]\boldsymbol{v}^d[n]$. Similarly, $\boldsymbol{c}[n]$ can be expressed as
\begin{equation}
    \boldsymbol{c}[n] = \left[|c_1[n]|e^{j\psi_1[n]}, |c_2[n]|e^{j\psi_2[n]}, \cdots, |c_{M+1}[n]|e^{j\psi_{M+1}[n]}\right],
\end{equation}
where $|c_k[n]|$ and $\psi_k[n]$ are the magnitude and phase angle of the $k$th element of $\boldsymbol{c}[n]$ in time slot $n$, respectively. Hence, $\Delta\boldsymbol{h}_{E1}^H[n]\mathbf{H}_{E1}[n]\boldsymbol{v}^d[n]$ can be given by
\begin{align}\label{eq.1}
    &\Delta\boldsymbol{h}_{E1}^H[n]\mathbf{H}_{E1}[n]\boldsymbol{v}^d[n] = \Delta\boldsymbol{h}_{E1}^H[n]\boldsymbol{c}[n]\nonumber\\
    & = |\Delta h_{E1,1}[n]||c_1[n]|e^{j\left(\psi_1[n]-\tau_1[n]\right)} + \cdots \nonumber\\
    &\quad+ |\Delta h_{E1,M+1}[n]||c_{M+1}[n]|e^{j\left(\psi_{M+1}[n]-\tau_{M+1}[n]\right)}.
\end{align}
It is known that the maximum  of $\left|\Delta\boldsymbol{h}_{E1}^H[n]\mathbf{H}_{E1}[n]\boldsymbol{v}^d[n]\right|$ can be obtained when all the items in the last step of \eqref{eq.1} can be coherently added. Hence, we have $\psi_1[n]-\tau_1[n] = \psi_2[n]-\tau_2[n] = \cdots = \psi_{M+1}[n]-\tau_{M+1}[n]$. Then, based on the constraints in \eqref{opterror.1.c}, it is not difficult to show that the optimal $\tau_k[n]$, denoted as $\tau_k^{op}[n]$, is given by
\begin{equation}
    \tau_k^{op}[n] = \psi_k[n] - \arg\left(\Bar{\boldsymbol{h}}_{E1}^H\mathbf{H}_{E1}[n]\boldsymbol{v}^d[n]\right).
\end{equation}
As such, problem \eqref{opterror.1} can be transformed into
\begin{subequations}\label{opterror.2}
\begin{align}
    &\max\limits_{\boldsymbol{m}_1[n]} ~ \left|\boldsymbol{m}_1^T[n]\boldsymbol{m}_2[n] \right|^2\label{opterror.2.a}\\
    &~~\textrm{s.t.} ~~~ \lVert\boldsymbol{m}_1[n] \rVert \leq \epsilon_1,\label{opterror.2.b}
\end{align}
\end{subequations}
where 
$$\boldsymbol{m}_1[n] = \left[|\Delta h_{E1,1}[n]|, |\Delta h_{E1,2}[n]|, \cdots, |\Delta h_{E1,M+1}[n]|\right]^T$$ and $$\boldsymbol{m}_2[n] = \left[|c_1[n]|, |c_2[n]|, \cdots, |c_{M+1}[n]|\right].$$ It is not difficult to show that the optimal $\boldsymbol{m}_1[n]$, denoted as $\boldsymbol{m}_1^{op}[n]$ is given by
\begin{equation}
    \boldsymbol{m}_1^{op}[n] = \frac{\epsilon_1}{\lVert\boldsymbol{m}_2[n]\rVert}\boldsymbol{m}_2[n].
\end{equation}
Therefore, the optimal $\Delta\boldsymbol{h}_{E1}[n]$, denoted as $\Delta\boldsymbol{h}_{E1}^{op}[n]$, is
\begin{equation}\label{opthe1.1}
    \Delta\boldsymbol{h}_{E1}^{op}[n] = {\rm diag}\left(\left[e^{j\tau_1^{op}[n]}, e^{j\tau_2^{op}[n]}, \cdots, e^{j\tau_{M+1}^{op}[n]}\right]\right)\boldsymbol{m}_1^{op}[n].
\end{equation}$\Delta\boldsymbol{h}_{E2}^{op}[n]$ can also be obtained by using the above solution. With $\Delta\boldsymbol{h}_{E1}^{op}[n]$ and $\Delta\boldsymbol{h}_{E2}^{op}[n]$, problem \eqref{optpower.1} can be rewritten as
\begin{align}\label{optpower.2}
    & \max\limits_{\mathbf{p}, \mathbf{g}} \quad \frac{1}{N}\sum\limits_{n=1}^N\left[ w\Tilde{R}_{down}^{power}[n] +  \left(1-w\right) \Tilde{R}_{up}^{power}[n]\right]\\
    & ~~\textrm{s.t.} ~~~~ \eqref{power.1}, \eqref{power_g.1},\nonumber
\end{align}
where 
$$\Tilde{R}_{down}^{power}[n]=\log_2\left(1 + p[n]a_1[n]\right)-\log_2\left(1 + p[n]b_1[n]\right),$$
$$\Tilde{R}_{up}^{power}[n]=\log_2\left(1 + g[n]a_1[n]\right)-\log_2\left(1 + g[n]b_2[n]\right),$$
$$a_1[n] = \frac{\left|\boldsymbol{h}_{G1}^H\mathbf{H}_{G1}[n]\boldsymbol{v}^d[n]\right|^2}{\sigma^2},$$
$$b_1[n] =\frac{\left|\left(\boldsymbol{h}_{E1}^{op}[n]\right)^H\mathbf{H}_{E1}[n]\boldsymbol{v}^d[n]\right|^2}{\sigma^2},$$
$$a_2[n] = \frac{\left|\boldsymbol{h}_{G2}^H[n]\mathbf{H}_{G2}[n]\boldsymbol{v}^u[n]\right|^2}{\sigma^2},$$
$$b_2[n] =\frac{\left|\left(\boldsymbol{h}_{E2}^{op}[n]\right)^H\mathbf{H}_{E2}\boldsymbol{v}^u[n]\right|^2}{\sigma^2}$$
$\boldsymbol{h}_{E1}^{op}[n] = \Bar{\boldsymbol{h}}_{E1} + \Delta\boldsymbol{h}_{E1}^{op}[n]$, and $\boldsymbol{h}_{E2}^{op}[n] = \Bar{\boldsymbol{h}}_{E2} + \Delta\boldsymbol{h}_{E2}^{op}[n]$. Similar to sub-problem 1 in \cite{zhang2019}, the optimal solution of \eqref{optpower.2} is given by
\begin{subequations}\label{solution_p.1}
\begin{align}
    p^{op}[n]=\begin{cases}
    \min\left(\left[\Tilde{p}[n]\right]^+, P_{peak}\right), &a_1[n] > b_1[n]\\
    0, &a_1[n] \leq b_1[n]
    \end{cases},\label{solution_p.1.a}\\
    g^{op}[n]=\begin{cases}
    \min\left(\left[\Tilde{g}[n]\right]^+, G_{peak}\right), &a_2[n] > b_2[n]\\
    0, &a_2[n] \leq b_2[n]
    \end{cases},\label{solution_p.1.b}
\end{align}
\end{subequations}
where
\begin{subequations}\label{tbpower.1}
\begin{align}
    \Tilde{p}[n]&=\sqrt{\left(\frac{1}{2b_1[n]}-\frac{1}{2a_1[n]}\right)^2+\frac{1}{\varpi_1\ln{2}}\left(\frac{1}{b_1[n]}-\frac{1}{a_1[n]}\right)}\nonumber\\
    &\quad-\frac{1}{2b_1[n]}-\frac{1}{2a_1[n]},\label{tbpower.1.a}\\
    \Tilde{g}[n]&=\sqrt{\left(\frac{1}{2b_2[n]}-\frac{1}{2a_2[n]}\right)^2+\frac{1}{\varpi_2\ln{2}}\left(\frac{1}{b_2[n]}-\frac{1}{a_2[n]}\right)}\nonumber\\
    &\quad-\frac{1}{2b_2[n]}-\frac{1}{2a_2[n]}.\label{tbpower.1.b}
\end{align}
\end{subequations}
Note that $\varpi_1 \ge 0$ and $\varpi_2 \ge 0$ in \eqref{tbpower.1} can be obtained via a one-dimensional bisection search, which guarantees that the constraints in \eqref{power.1.a} and \eqref{power_g.1.a} are fulfilled when $p^{op}[n]$ and $g^{op}[n]$ are attained, respectively.

\subsection{Solution to Sub-Problem 2}
For any given $\mathbf{p}$, $\mathbf{g}$, and $\mathbf{Q}$, with the aid of the slack variables $\boldsymbol{\xi_1}=\{\xi_1[n]\}_{n=1}^N$ and $\boldsymbol{\xi_2}=\{\xi_2[n]\}_{n=1}^N$, sub-problem 2 can be expressed as
\begin{subequations}\label{optphi.1}
\begin{align}
    & \max\limits_{\boldsymbol{v}^d[n], \boldsymbol{v}^u[n],
    \atop
    \xi_1[n], \xi_2[n]} ~ \frac{1}{N}\sum\limits_{n=1}^N \left[ w R_{down}^{phi}[n] +  \left(1-w\right) R_{up}^{phi}[n]\right]\label{optphi_g.1.a}\\
    &\quad~\textrm{s.t.} \max\limits_{\Delta\boldsymbol{h}_{E1}[n]\in\Omega_1} \frac{p[n]}{\sigma^2}\left|\boldsymbol{h}_{E1}^H[n]\mathbf{H}_{E1}[n]\boldsymbol{v}^d[n]\right|^2 \leq \xi_1[n], \forall n,\label{optphi.1.b}\\
    &\quad\quad~~\max\limits_{\Delta\boldsymbol{h}_{E2}[n]\in\Omega_2} \frac{g[n]}{\sigma^2}\left|\boldsymbol{h}_{E2}^H[n]\mathbf{H}_{E2}\boldsymbol{v}^u[n]\right|^2 \leq \xi_2[n], \forall n,\label{optphi.1.c}\\
    &\quad\quad\quad\quad|v_i^d[n]|, |v_i^u[n]|=1, \forall n,i,\label{optphi.1.d}
\end{align}
\end{subequations}where 
\begin{equation}
    \begin{split}
        R_{down}^{phi}[n]&=\log_2\left(1 + \frac{p[n]}{\sigma^2} \left|\boldsymbol{h}_{G1}^H\mathbf{H}_{G1}[n]\boldsymbol{v}^d[n]\right|^2 \right)\\
    &\quad- \log_2\left(1 + \xi_1[n] \right)\nonumber
    \end{split}
\end{equation}
and
\begin{equation}
    \begin{split}
        R_{up}^{phi}[n]&=\log_2\left(1 + \frac{g[n]}{\sigma^2} \left|\boldsymbol{h}_{G2}^H[n]\mathbf{H}_{G2}[n]\boldsymbol{v}^u[n]\right|^2 \right)\\
    &\quad - \log_2\left(1 + \xi_2[n] \right).\nonumber
    \end{split}
\end{equation}
It is difficult to solve problem \eqref{optphi.1}, since the constraints in \eqref{optphi.1.b} and \eqref{optphi.1.c} involve infinitely many inequality constraints. To overcome this difficulty, we first substitute \eqref{uncertainty.1.a} and \eqref{uncertainty.1.b} into \eqref{optphi.1.b} and \eqref{optphi.1.c}, respectively, and obtain
\begin{subequations}\label{uncertainty.2}
    \begin{align}
        &\Delta\boldsymbol{h}_{E1}^H[n]\Delta\boldsymbol{h}_{E1}[n] - \epsilon_1^2 \leq 0, \forall n\label{uncertainty.2.a}\\
        &\Delta\boldsymbol{h}_{E2}^H[n]\Delta\boldsymbol{h}_{E2}[n] - \epsilon_2^2 \leq 0, \forall n\label{uncertainty.2.b}\\
        &\frac{p[n]}{\sigma^2}\boldsymbol{h}_{E1}^H[n]\mathbf{H}_{E1}[n]\boldsymbol{V}^d[n]\mathbf{H}_{E1}^H[n]\boldsymbol{h}_{E1}[n] -\xi_1[n] \leq 0, \forall n,\label{uncertainty.2.c}\\
        &\frac{g[n]}{\sigma^2}\boldsymbol{h}_{E2}^H[n]\mathbf{H}_{E2}\boldsymbol{V}^u[n]\mathbf{H}_{E2}^H\boldsymbol{h}_{E2}[n]-\xi_2[n] \leq 0, \forall n,\label{uncertainty.2.d}
    \end{align}
\end{subequations}where $\boldsymbol{V}^d[n]=\boldsymbol{v}^d[n]{\boldsymbol{v}^d[n]}^H$ and $\boldsymbol{V}^u[n]=\boldsymbol{v}^u[n]{\boldsymbol{v}^u[n]}^H$. The ranks of $\boldsymbol{V}^d[n]$ and $\boldsymbol{V}^u[n]$ are one. Then, we transform the constraints in \eqref{optphi.1.b} and \eqref{optphi.1.c} into linear matrix inequalities (LMIs) by using the following lemma. 
\begin{lem}
    ($\mathcal{S}$-Procedure \cite{schober2014}:) Let a function $f_m(\boldsymbol{x}), m\in\{1,2\}, \boldsymbol{x} \in \mathbb{C}^{N \times 1}$, be defined as 
    \begin{equation}
        f_m(\boldsymbol{x}) = \boldsymbol{x}^H\mathbf{B}_m\boldsymbol{x}+2{\rm Re}\{\boldsymbol{b}_m^H\boldsymbol{x}\} + b_m,
    \end{equation}
    where $\mathbf{B}_m \in \mathbb{H}^N$, $\boldsymbol{b}_m \in \mathbb{C}^{N \times 1}$, and $b_m \in \mathbb{R}^{1\times1}$. Then, the implication $f_1(\boldsymbol{x})\leq0 \Rightarrow{f_2(\boldsymbol{x})\leq0}$ hold if and only if there exists a $\delta\ge0$ such that
    \begin{equation}
        \delta\begin{bmatrix}\mathbf{B}_1 \quad \boldsymbol{b}_1 \\ \boldsymbol{b}_1^H \quad b_1\end{bmatrix} - \begin{bmatrix}\mathbf{B}_2 \quad \boldsymbol{b}_2 \\ \boldsymbol{b}_2^H \quad b_2\end{bmatrix} \succeq 0,
    \end{equation}
    provided that there exists a point $\hat{\boldsymbol{x}}$ such that $f_m(\hat{\boldsymbol{x}})<0$.
\end{lem}

\newcounter{TempEqCnt}
\setcounter{TempEqCnt}{\value{equation}} 
\setcounter{equation}{42} 
\begin{figure*}[ht] 
	\begin{equation}\label{U2}
	\boldsymbol{U}_2[n]=\frac{p[n]}{\sigma^2}\begin{bmatrix}\mathbf{H}_{E1}[n]\boldsymbol{V}^d[n]\mathbf{H}_{E1}^H[n] & \mathbf{H}_{E1}[n]\boldsymbol{V}^d[n]\mathbf{H}_{E1}^H[n]\Bar{\boldsymbol{h}}_{E1}\\
    \Bar{\boldsymbol{h}}_{E1}^H \mathbf{H}_{E1}[n]\boldsymbol{V}^d[n]\mathbf{H}_{E1}^H[n] & \Bar{\boldsymbol{h}}_{E1}^H \mathbf{H}_{E1}[n]\boldsymbol{V}^d[n]\mathbf{H}_{E1}^H[n]\Bar{\boldsymbol{h}}_{E1}\end{bmatrix}
	\end{equation}
	\begin{equation}\label{U4}
	\boldsymbol{U}_4[n]=\frac{g[n]}{\sigma^2}\begin{bmatrix}\mathbf{H}_{E2}\boldsymbol{V}^u[n]\mathbf{H}_{E2}^H & \mathbf{H}_{E2}\boldsymbol{V}^u[n]\mathbf{H}_{E2}^H\Bar{\boldsymbol{h}}_{E2}\\
    \Bar{\boldsymbol{h}}_{E2}^H \mathbf{H}_{E2}\boldsymbol{V}^u[n]\mathbf{H}_{E2}^H & \Bar{\boldsymbol{h}}_{E2}^H \mathbf{H}_{E2}\boldsymbol{V}^u[n]\mathbf{H}_{E2}^H\Bar{\boldsymbol{h}}_{E2}\end{bmatrix}
	\end{equation}
	\hrulefill
\end{figure*}
\setcounter{equation}{\value{TempEqCnt}}
Using Lemma 1, the following implications can be obtained:  $\eqref{uncertainty.2.a}\Rightarrow{\eqref{uncertainty.2.c}}$ and $\eqref{uncertainty.2.b}\Rightarrow{\eqref{uncertainty.2.d}}$ holds if and only if there exist $\eta_1[n] \ge 0$ and $\eta_2[n] \ge 0$ such that
\begin{subequations}\label{LMI.2}
\begin{align}
    &\boldsymbol{U}_1[n] - \boldsymbol{U}_2[n] \succeq 0,\label{LMI.2.a}\\
    &\boldsymbol{U}_3[n] - \boldsymbol{U}_4[n] \succeq 0,\label{LMI.2.b}
\end{align}
\end{subequations}
where 
\begin{align}
    \boldsymbol{U}_1[n]&=\begin{bmatrix}\eta_1[n]\mathbf{I}_{M+1} & 0\\
    0 & -\eta_1[n]\epsilon_1^2+\xi_1[n]\end{bmatrix},\\
    \boldsymbol{U}_3[n]&=\begin{bmatrix}\eta_2[n]\mathbf{I}_{M+1} & 0\\
    0 & -\eta_2[n]\epsilon_2^2+\xi_2[n]\end{bmatrix},
\end{align}
$\mathbf{I}_{M+1}$ denotes the $\left(M+1\right) \times \left(M+1\right)$ identity matrix, and $\boldsymbol{U}_2[n]$ and $\boldsymbol{U}_4[n]$ are given in \eqref{U2} and \eqref{U4}, respectively. Since the unit-modulus constraints in \eqref{optphi.1.d} are non-convex, we apply the SDR method to relax the constraints. We have 
\begin{align}
    \left|\boldsymbol{h}_{G1}^H\mathbf{H}_{G1}[n]\boldsymbol{v}^d[n]\right|^2 &= \boldsymbol{h}_{G1}^H\mathbf{H}_{G1}[n]\boldsymbol{V}^d[n]\mathbf{H}_{G1}^H[n]\boldsymbol{h}_{G1}\nonumber\\
    &={\rm Tr}\left(\boldsymbol{V}^d[n]\mathbf{A}_1[n] \right)\nonumber\\
    \left|\boldsymbol{h}_{G2}^H[n]\mathbf{H}_{G2}[n]\boldsymbol{v}^u[n]\right|^2 &= \boldsymbol{h}_{G2}^H[n]\mathbf{H}_{G2}[n]\boldsymbol{V}^u[n]\mathbf{H}_{G2}^H[n]\boldsymbol{h}_{G2}[n]\nonumber\\
    &={\rm Tr}\left(\boldsymbol{V}^u[n]\mathbf{A}_2[n] \right),\nonumber
\end{align}
where 
$$\mathbf{A}_1[n] = \mathbf{H}_{G1}^H[n]\boldsymbol{h}_{G1}\boldsymbol{h}_{G1}^H\mathbf{H}_{G1}[n],$$
$$\mathbf{A}_2[n] = \mathbf{H}_{G2}^H[n]\boldsymbol{h}_{G2}[n]\boldsymbol{h}_{G2}^H[n]\mathbf{H}_{G2}[n],$$
and ${\rm Tr}\left(\boldsymbol{X}\right)$ denotes the trace of $\boldsymbol{X}$. Thus, problem \eqref{optphi.1} can be reformulated as
\setcounter{equation}{44} 
\begin{subequations}\label{optphi.2}
\begin{align}
    & \max\limits_{\boldsymbol{V}^d[n], \boldsymbol{V}^u[n], \xi_1[n],
    \atop
    \xi_2[n], \eta_1[n], \eta_2[n]} ~ \frac{1}{N}\sum\limits_{n=1}^N \left[ w\Tilde{R}_{down}^{phi}[n] + \left(1 - w\right) \Tilde{R}_{up}^{phi}[n]\right]\label{optphi.2.a}\\
    & \quad~~~~\textrm{s.t.} \quad\quad~~\eta_1[n], \eta_2[n]\ge0, \forall n, \label{optphi.2.b}\\
    & ~~ \quad\quad\quad\quad\quad ~~ \boldsymbol{V}^d[n], \boldsymbol{V}^u[n]\succeq 0, \forall n, \label{optphi.2.c} \\
    & ~~ \quad\quad\quad\quad\quad ~~ \boldsymbol{V}^d_{r,r}[n], \boldsymbol{V}^u_{r,r}[n]=1, r=1, \cdots, M+1, \forall n,\label{optphi.2.d}\\
    & ~~ \quad\quad\quad\quad\quad ~~ \eqref{LMI.2},\nonumber
\end{align}
\end{subequations}
where 
\begin{align}
    \Tilde{R}_{down}^{phi}[n]&=\log_2\left(1 + \frac{p[n]}{\sigma^2} {\rm Tr}\left(\boldsymbol{V}^d[n]\mathbf{A}_1[n] \right) \right)\nonumber\\
    &\quad-\log_2\left(1 + \xi_1[n] \right),\nonumber\\
    \Tilde{R}_{up}^{phi}[n]&=\log_2\left(1 + \frac{g[n]}{\sigma^2} {\rm Tr}\left(\boldsymbol{V}^u[n]\mathbf{A}_2[n] \right) \right)\nonumber\\
    &\quad- \log_2\left(1 + \xi_2[n] \right),\nonumber
\end{align}
and $\boldsymbol{V}^d_{r,r}[n]$ and $\boldsymbol{V}^u_{r,r}[n]$ denote the $\left(r, r\right)$th element of $\boldsymbol{V}^d[n]$ and $\boldsymbol{V}^u[n]$, respectively. It is still difficult to obtain the optimal solution of problem \eqref{optphi.2}, since $-\log_2\left(1 + \xi_1[n] \right)$ and $-\log_2\left(1 + \xi_2[n] \right)$ are not concave with respect to $\xi_1[n]$ and $\xi_1[n]$, respectively. Nevertheless, it is known that the first-order Taylor expansion of a concave function is its global over-estimator and that of a convex function is its global under-estimator. Therefore, we apply the SCA method to solve problem \eqref{optphi.2}. The first-order Taylor expansions of $\log_2\left(1 + \xi_1[n] \right)$ and $\log_2\left(1 + \xi_2[n] \right)$ at the given points $\boldsymbol{\xi}_{1, 0}=\{\xi_{1, 0}[n]\}_{n=1}^N$ and $\boldsymbol{\xi}_{2, 0}=\{\xi_{2,0}[n]\}_{n=1}^N$ can be respectively expressed as
\begin{align}
    \log_2\left(1 + \xi_1[n] \right) &\leq \log_2\left(1 + \xi_{1, 0}[n] \right) \nonumber\\
    &+ \frac{1}{\ln{2}\left(1 + \xi_{1, 0}[n] \right)}\left(\xi_1[n] -\xi_{1, 0}[n] \right),\\
    \log_2\left(1 + \xi_2[n] \right) &\leq \log_2\left(1 + \xi_{2, 0}[n] \right) \nonumber\\
    &+ \frac{1}{\ln{2}\left(1 + \xi_{2, 0}[n] \right)}\left(\xi_2[n] -\xi_{2, 0}[n] \right).
\end{align}
Then, problem \eqref{optphi.2} can be approximated as
\begin{align}\label{optphi.3}
    & \max\limits_{\boldsymbol{V}^d[n], \boldsymbol{V}^u[n], \xi_1[n],
    \atop
    \xi_2[n], \eta_1[n], \eta_2[n]} ~ \frac{1}{N}\sum\limits_{n=1}^N \left[ w \Hat{R}_{down}^{phi}[n] + \left(1 - w\right) \Hat{R}_{up}^{phi}[n]\right]\\
    & \quad\quad~\textrm{s.t.} \quad\quad~~\eqref{LMI.2}, \eqref{optphi.2.b}, \eqref{optphi.2.c}, \eqref{optphi.2.d},\nonumber
\end{align}
where
\begin{equation}
    \begin{split}
       \Hat{R}_{down}^{phi}[n]&=\log_2\left(1 + \frac{p[n]}{\sigma^2} {\rm Tr}\left(\boldsymbol{V}^d[n]\mathbf{A}_1[n] \right) \right)\\
    &\quad-\frac{\xi_1[n]}{\ln{2}\left(1 + \xi_{1, 0}[n] \right)}\nonumber
    \end{split}
\end{equation}
and
\begin{equation}
    \begin{split}
       \Hat{R}_{up}^{phi}[n]&=\log_2\left(1 + \frac{g[n]}{\sigma^2} {\rm Tr}\left(\boldsymbol{V}^u[n]\mathbf{A}_2[n] \right) \right)\\
    &\quad-\frac{\xi_2[n]}{\ln{2}\left(1 + \xi_{2, 0}[n] \right)}.\nonumber
    \end{split}
\end{equation}
It is observed that problem \eqref{optphi.3} is a convex optimization problem, and thus can be solved efficiently by using standard solvers, such as the CVX. However, we emphasize that a rank-one solution may not be obtained. Hence, we use the Gaussian randomization method \cite{wu2019intelligent} to recover $\boldsymbol{v}^d[n]$ and $\boldsymbol{v}^u[n]$ from $\boldsymbol{V}^d[n]$ and $\boldsymbol{V}^u[n]$, respectively, which is similar to that in \cite{wu2019intelligent} and thus omitted here for brevity.

\subsection{Solution to Sub-Problem 3}
For any given $\boldsymbol{\Phi}_d$, $\boldsymbol{\Phi}_u$, $\mathbf{p}$, and $\mathbf{g}$, we can express sub-problem 3 as 
\begin{align}\label{opttraj_a.1}
     & \max\limits_{\mathbf{Q}} ~ \frac{1}{N}\sum\limits_{n=1}^N
     \biggl[ \left(1-w\right) \log_2\biggl(1 + \frac{g[n]}{\sigma^2} \left|\boldsymbol{h}_{G2}^H[n]\mathbf{H}_{G2}[n]\boldsymbol{v}^u[n]\right|^2\biggr)\nonumber\\
     &\quad\quad\quad\quad\quad~~+w R_{down}^{traj}[n]\biggr]
     \\
     & ~~\textrm{s.t.} ~~ \eqref{mobility.1}, \nonumber
\end{align}
where 
\begin{align}
    R_{down}^{traj}[n] &= \log_2\left(1 + \frac{p[n]}{\sigma^2} \left|\boldsymbol{h}_{G1}^H\mathbf{H}_{G1}[n]\boldsymbol{v}^d[n]\right|^2\right) \nonumber\\
    &\quad- \log_2\left(1+\max\limits_{\Delta\boldsymbol{h}_{E1}\in\Omega_1}\frac{p[n]}{\sigma^2}\left|\boldsymbol{h}_{E1}^H\mathbf{H}_{E1}[n]\boldsymbol{v}^d[n]\right|^2\right).\nonumber
\end{align}
\newcounter{TempEqCnt1}
\setcounter{TempEqCnt1}{\value{equation}} 
\setcounter{equation}{53} 
\begin{figure*}[ht] 
	\begin{equation}\label{hqg}
	\mathbf{H}_{QG}[n] = \left[h_{U\!G}^H, \sqrt{(d_{RG})^{-\alpha}} (\boldsymbol{h}_{U\!R}^{(j-1)}[n])^H\boldsymbol{\Theta}_d^H[n] \boldsymbol{h}_{RG} \right]^H\left[h_{U\!G}^H, \sqrt{(d_{RG})^{-\alpha}} (\boldsymbol{h}_{U\!R}^{(j-1)}[n])^H\boldsymbol{\Theta}_d^H[n] \boldsymbol{h}_{RG} \right]
	\end{equation}
	\begin{equation}\label{hgq}
	\mathbf{H}_{GQ}[n] = \left[h_{GU}^H, \sqrt{(d_{RG})^{-\alpha}} \boldsymbol{h}_{G\!R}^H\boldsymbol{\Theta}_u^H[n] \boldsymbol{h}_{RU}^{(j-1)}[n] \right]^H\left[h_{GU}^H, \sqrt{(d_{RG})^{-\alpha}} \boldsymbol{h}_{G\!R}^H\boldsymbol{\Theta}_u^H[n] \boldsymbol{h}_{RU}^{(j-1)}[n] \right]
	\end{equation}
	\begin{equation}\label{hqe}
	\mathbf{H}_{QE}[n] = \left[\begin{matrix}h_{U\!E}^{op}[n] \\ \sqrt{(d_{RE})^{-\alpha}}\left(\boldsymbol{h}_{RE}^{op}[n]\right)^H \boldsymbol{\Theta}_d[n] \boldsymbol{h}_{U\!R}^{(j-1)}[n] \end{matrix}\right]\left[\begin{matrix}h_{U\!E}^{op}[n] \\ \sqrt{(d_{RE})^{-\alpha}}\left(\boldsymbol{h}_{RE}^{op}[n]\right)^H \boldsymbol{\Theta}_d[n] \boldsymbol{h}_{U\!R}^{(j-1)}[n] \end{matrix}\right]^H
	\end{equation}
	\hrulefill
\end{figure*}
\setcounter{equation}{\value{TempEqCnt1}}
In particular, $R_{G\!E}[n]$ is not relevant to the UAV trajectory, and so we omit it in problem \eqref{opttraj_a.1}. It is challenging to cope with the infinitely many $\Delta\boldsymbol{h}_{E1}$. However, we note that the worst case of the objective function is obtained when the UAV trajectory is given, and the UAV trajectory can be optimized when the worst case of the wiretap channels, i.e., $\boldsymbol{h}_{E1}^{op}[n]$, is given. Hence, we utilize the UAV trajectory of the $(j-1)$th iteration to calculate the worst case setup for the wiretap channels $\boldsymbol{h}_{E1}^{op}[n]$ in the $j$th iteration, and this is obtained by using a procedure similar to \eqref{opthe1.1}. Besides, from \eqref{LoScomponent.1}-(8), it is worth noting that not only $L_{U\!G}[n]$, $L_{U\!E}[n]$, $L_{U\!RG}[n]$, and $L_{U\!RE}[n]$ but also $\boldsymbol{h}_{U\!R}^{\rm{LoS}}[n]$ is relevant to the UAV trajectory. However, from the structure of $\boldsymbol{h}_{U\!R}^{\rm{LoS}}[n]$ in \eqref{LoScomponent.1}, it is observed that $\boldsymbol{h}_{U\!R}^{\rm{LoS}}[n]$ is complex and non-linear with respect to the UAV trajectory variables, which makes the UAV trajectory design intractable. To handle such difficulty, we use the UAV trajectory of the $(j-1)$th iteration to obtain an approximate $\boldsymbol{h}_{U\!R}^{\rm{LoS}}[n]$ in the $j$th iteration. Similarly, the LoS component of $\boldsymbol{h}_{RU}[n]$ in the $j$th iteration is also designed by using the UAV trajectory of the $(j-1)$th iteration. \textcolor{black}{Then, we can rewritten problem \eqref{opttraj_a.1} as 
\begin{align}\label{traj.1}
    & \max\limits_{\mathbf{Q}} ~ \frac{1}{N}\sum\limits_{n=1}^N
     \bigl[  \left(1-w\right) \log_2\left(1 + \rho\gamma_1[n]\boldsymbol{h}_{ue}^T[n]\mathbf{H}_{GQ}[n]\boldsymbol{h}_{ue}[n]\right)\nonumber\\
     &\quad\quad\quad\quad\quad~~+w \Dot{R}_{down}^{traj}[n]\bigr]
     \\
     & ~~\textrm{s.t.} ~~ (1), \nonumber
\end{align}
where 
\begin{align}
    \Dot{R}_{down}^{traj}[n] &= \log_2\left(1 + \rho\gamma_0[n] \boldsymbol{h}_{ue}^T[n]\mathbf{H}_{QG}[n]\boldsymbol{h}_{ue}[n]\right) \nonumber\\
    &\quad- \log_2\left(1+\rho\gamma_0[n]\boldsymbol{h}_{st}^T[n]\mathbf{H}_{QE}[n]\boldsymbol{h}_{st}[n]\right),\\
    \boldsymbol{h}_{ue}[n] &= \left[ \sqrt{(d_{U\!G}[n])^{-\kappa}}, \sqrt{(d_{U\!R}[n])^{-\alpha}}\right]^T,\\
    \boldsymbol{h}_{st}[n] &= \left[ \sqrt{(d_{U\!E}[n])^{-\kappa}}, \sqrt{(d_{U\!R}[n])^{-\alpha}}\right]^T,
\end{align}
$\mathbf{H}_{QG}[n]$, $\mathbf{H}_{GQ}[n]$, and $\mathbf{H}_{QE}[n]$ are given in \eqref{hqg}, \eqref{hgq}, and \eqref{hqe}, respectively, shown at the top of the next page; $\gamma_0[n]=p[n]/\sigma^2$, $\gamma_1[n]=g[n]/\sigma^2$, and $\boldsymbol{h}_{U\!R}^{(j-1)}[n]$ and $\boldsymbol{h}_{RU}^{(j-1)}[n]$ are the designed $\boldsymbol{h}_{U\!R}[n]$ and $\boldsymbol{h}_{RU}[n]$, respectively, by using the UAV trajectory of the $(j-1)$th iteration. Note that problem \eqref{traj.1} is not a convex problem due to the non-concave objective function with respect to the UAV trajectory $\mathbf{Q}$. By introducing the slack variables $\mathbf{u}=\left\{ u[n]\right\} _{n=1}^{N}$, $\mathbf{e}=\left\{ e[n]\right\} _{n=1}^{N}$, $\mathbf{s}=\left\{ s[n]\right\} _{n=1}^{N}$, $\mathbf{t}=\left\{ t[n]\right\} _{n=1}^{N}$, $\boldsymbol{\zeta}=\left\{ \zeta[n]\right\} _{n=1}^{N}$, $\mathbf{r}_d=\left\{ r_d[n]\right\} _{n=1}^{N}$, and $\mathbf{r}_u=\left\{ r_u[n]\right\} _{n=1}^{N}$, we transform problem \eqref{traj.1} into the following problem,
\setcounter{equation}{56} 
\begin{subequations}\label{opttraj.1}
\begin{align}
     & \max\limits_{\mathbf{Q}, \mathbf{u}, \mathbf{e}, \mathbf{s},
     \atop
     \mathbf{t}, \boldsymbol{\zeta}, \mathbf{r}_d, \mathbf{r}_u} ~ \frac{1}{N}\sum\limits_{n=1}^N\left[ w\Tilde{R}_{down}^{traj}[n] + \left(1-w\right) \log_2\left(1+\rho\gamma_1[n]r_u[n]\right)\right]
     \label{opttraj.1.a}\\
    & \quad~\textrm{s.t.} \quad \sqrt{(d_{U\!G}[n])^{-\kappa}} \ge u[n], \forall n, \label{opttraj.1.b}\\
     & \quad\quad\quad~~ \sqrt{(d_{U\!R}[n])^{-\alpha}} \ge e[n], \forall n, \label{opttraj.1.c}\\
      & \quad\quad\quad~~ \sqrt{(d_{U\!E}[n])^{-\kappa}} \leq s[n], \forall n, \label{opttraj.1.d}\\
     & \quad\quad\quad~~ \sqrt{(d_{U\!R}[n])^{-\alpha}} \leq t[n], \forall n, \label{opttraj.1.e}\\
    & \quad\quad\quad~~ \rho\gamma_0[n]\Tilde{\boldsymbol{h}}_{st}^T[n]\mathbf{H}_{QE}[n]\Tilde{\boldsymbol{h}}_{st}[n] \leq \zeta[n], \forall n, \label{opttraj.1.f}\\
     & \quad\quad\quad~~ \Tilde{\boldsymbol{h}}_{ue}^T[n]\mathbf{H}_{QG}[n]\Tilde{\boldsymbol{h}}_{ue}[n]\ge r_d[n], \forall n,\label{opttraj.1.g}\\
     & \quad\quad\quad~~ \Tilde{\boldsymbol{h}}_{ue}^T[n]\mathbf{H}_{GQ}[n]\Tilde{\boldsymbol{h}}_{ue}[n]\ge r_u[n], \forall n,\label{opttraj.1.h}\\
     &\quad\quad\quad~~ s[n] \leq \sqrt{(z_U-z_E)^{-\kappa}}, ~t[n] \leq \sqrt{(z_U-z_R)^{-\alpha}}, \forall n,\label{opttraj.1.i}\\
     & \quad\quad \quad~~ (1), \nonumber
 \end{align}
\end{subequations}
where $\Tilde{R}_{down}^{traj}[n] = \log_2\left(1+\rho\gamma_0[n]r_d[n]\right) - \log_2\left(1+\zeta[n]\right)$, $\Tilde{\boldsymbol{h}}_{ue}[n] = \left[ u[n], e[n] \right]^T$, and $\Tilde{\boldsymbol{h}}_{st}[n] = \left[ s[n], t[n]\right]^T$. The constraint \eqref{opttraj.1.i} gives the upper bounds of $s[n]$ and $t[n]$. This is because the maximums of $\sqrt{(d_{U\!E}[n])^{-\kappa}}$ and $\sqrt{(d_{U\!R}[n])^{-\alpha}}$ 
are given by $\sqrt{(z_U-z_E)^{-\kappa}}$ and $\sqrt{(z_U-z_R)^{-\alpha}}$ when the UAV hovers above the eavesdropper and the RIS, respectively. Hence, we have $s[n] \leq \sqrt{(z_U-z_E)^{-\kappa}}$ and $t[n] \leq \sqrt{(z_U-z_R)^{-\alpha}}$. In order to facilitate the subsequent derivations, we unfold \eqref{opttraj.1.b}-\eqref{opttraj.1.e} as follows,
\begin{subequations}\label{unfold.1}
\begin{align}
    &x^2[n]+x^2_G+y^2[n]+y^2_G-2x_G x[n]-2y_G y[n]\nonumber\\
    &+z_U^2-\left(u[n]\right)^{-\frac{4}{\kappa}}\leq0, \forall n, \label{unfold.1.a}\\
    &  x^2[n]+x^2_R+y^2[n]+y^2_R-2x_Rx[n]-2y_Ry[n]\nonumber\\
    &+(z_U-z_R)^2-\left(e[n]\right)^{-\frac{4}{\alpha}}\leq0, \forall n, \label{unfold.1.b}\\
    &  \left(s[n]\right)^{-\frac{4}{\kappa}}-x^2[n]-x^2_E-y^2[n]-y^2_E+2x_E x[n]\nonumber\\
    &+2y_E y[n]-z_U^2 \leq 0, \forall n, \label{unfold.1.c}\\
    &  \left(t[n]\right)^{-\frac{4}{\alpha}}-x^2[n]-x^2_R-y^2[n]-y^2_R+2x_Rx[n]\nonumber\\
    &+2y_Ry[n]-(z_U-z_R)^2 \leq 0, \forall n. \label{unfold.1.d}
\end{align}
\end{subequations} 
It is observed that the constraints \eqref{unfold.1.a}-\eqref{unfold.1.d}} are non-convex feasible regions, and $- \log_2\left(1+\zeta[n]\right)$ is non-concave with respect to $\zeta[n]$. 
We use the SCA technique to address the non-convexity of these constraints. The first-order Taylor expansions of $-x^2[n]$, $-y^2[n]$, $\left(u[n]\right)^{-\frac{4}{\kappa}}$, $\left(e[n]\right)^{-\frac{4}{\alpha}}$, $\log_2\left(1 + \zeta[n] \right)$, $\Tilde{\boldsymbol{h}}_{ue}^T[n]\mathbf{H}_{QG}[n]\Tilde{\boldsymbol{h}}_{ue}[n]$, and $\Tilde{\boldsymbol{h}}_{ue}^T[n]\mathbf{H}_{GQ}[n]\Tilde{\boldsymbol{h}}_{ue}[n]$ at the given feasible points $\mathbf{x}_0=\left\{ x_0[n]\right\} _{n=1}^{N}$, $\mathbf{y}_0=\left\{y_0[n]\right\} _{n=1}^{N}$, $\mathbf{u}_0=\left\{ u_0[n]\right\} _{n=1}^{N}$, $\mathbf{e}_0=\left\{ e_0[n]\right\} _{n=1}^{N}$, $\boldsymbol{\zeta}_0=\left\{ \zeta_0[n]\right\} _{n=1}^{N}$, and $\mathbf{H}_{ue,0}=\left\{\Tilde{\boldsymbol{h}}_{ue,0}[n]\right\}_{n=1}^N$ are given by 
\textcolor{black}{\begin{subequations}\label{taylor.1}
\begin{align}
    &\Tilde{\boldsymbol{h}}_{ue}^T[n]\mathbf{H}_{QG}[n]\Tilde{\boldsymbol{h}}_{ue}[n] \ge -\Tilde{\boldsymbol{h}}_{ue,0}^T[n]\mathbf{H}_{QG}[n]\Tilde{\boldsymbol{h}}_{ue,0}[n] \nonumber\\
    &\quad\quad\quad\quad\quad\quad\quad\quad~+ 2\Re\left[\Tilde{\boldsymbol{h}}_{ue,0}^T[n]\mathbf{H}_{QG}[n]\Tilde{\boldsymbol{h}}_{ue}[n]\right],\label{taylor.1.a}\\
    &\Tilde{\boldsymbol{h}}_{ue}^T[n]\mathbf{H}_{GQ}[n]\Tilde{\boldsymbol{h}}_{ue}[n] \ge -\Tilde{\boldsymbol{h}}_{ue,0}^T[n]\mathbf{H}_{GQ}[n]\Tilde{\boldsymbol{h}}_{ue,0}[n]\nonumber\\
    &\quad\quad\quad\quad\quad\quad\quad\quad~+ 2\Re\left[\Tilde{\boldsymbol{h}}_{ue,0}^T[n]\mathbf{H}_{GQ}[n]\Tilde{\boldsymbol{h}}_{ue}[n]\right],\label{taylor.1.b}\\
    &\log_2\left(1 + \zeta[n] \right) \leq \log_2\left(1 + \zeta_0[n] \right) \nonumber\\
    &\quad\quad\quad\quad\quad\quad\quad~+ \frac{1}{\ln{2}\left(1 + \zeta_0[n] \right)}\left(\zeta[n] -\zeta_0[n] \right),\label{taylor.1.c}\\
    &\left(u[n]\right)^{-\frac{4}{\kappa}} \ge \left(u_0[n]\right)^{-\frac{4}{\kappa}} - \frac{4}{\kappa}\left(u_0[n]\right)^{-\frac{4}{\kappa}-1}\left(u[n]-u_0[n]\right),\label{taylor.1.d}\\
    &\left(e[n]\right)^{-\frac{4}{\alpha}} \ge \left(e_0[n]\right)^{-\frac{4}{\alpha}} - \frac{4}{\alpha}\left(e_0[n]\right)^{-\frac{4}{\alpha}-1}\left(e[n]-e_0[n]\right),\label{taylor.1.e}\\
    &\quad\quad\quad\quad\quad-x^2[n] \leq x_0^2[n] - 2x_0[n]x[n],\label{taylor.1.f}\\
    &\quad\quad\quad\quad\quad-y^2[n] \leq y_0^2[n] - 2y_0[n]y[n].\label{taylor.1.g}
\end{align}
\end{subequations}}

\begin{algorithm}[t] 
    \caption{Proposed algorithm for solving problem \eqref{optimal.1}} 
    \label{alg1} 
    \begin{algorithmic}[1] 
    \STATE \textbf{Initialization}: \\
    Set the initial feasible points $\Xi_0 = \{\mathbf{Q}^{(0)}, \boldsymbol{\Phi}_d^{(0)}, \boldsymbol{\Phi}_u^{(0)}, \mathbf{p}^{(0)}, \mathbf{g}^{(0)}, \mathbf{u}_0, \mathbf{e}_0, \boldsymbol{\xi}_{1,0}, \boldsymbol{\xi}_{2,0}, \boldsymbol{\zeta}_0\}$. Set iteration index $j = 0$ and $R_{sec}^{(0)}$.
    \STATE \textbf{repeat}
    \STATE \quad Set $j\gets j+1$;
    \STATE \quad With given $\mathbf{Q}^{(j-1)}$, $\mathbf{p}^{(j-1)}$, $\mathbf{g}^{(j-1)}$, $\boldsymbol{\Phi}_d^{(j-1)}$, $\boldsymbol{\Phi}_u^{(j-1)}$, $\mathbf{u}_0$, $\mathbf{e}_0$, and $\boldsymbol{\zeta}_0$, update $\mathbf{Q}^{(j)}$, $\mathbf{u}^{(j)}$, $\mathbf{e}^{(j)}$, 
    
    \setlength{\parindent}{1em} and
    $\boldsymbol{\zeta}^{(j)}$ by solving problem \eqref{opttraj.2};
     
    \STATE \quad Set $\mathbf{u}_0 = \mathbf{u}^{(j)}$, $\mathbf{e}_0 = \mathbf{e}^{(j)}$, and $\boldsymbol{\zeta}_0 = \boldsymbol{\zeta}^{(j)}$;
    \STATE \quad With given $\mathbf{Q}^{(j)}$, $\mathbf{p}^{(j-1)}$, $\mathbf{g}^{(j-1)}$, $\boldsymbol{\xi}_{1,0}$, and $\boldsymbol{\xi}_{2,0}$, update $\boldsymbol{\Phi}_d^{(j)}$, $\boldsymbol{\Phi}_u^{(j)}$, $\boldsymbol{\xi}_{1}^{(j)}$, and $\boldsymbol{\xi}_{2}^{(j)}$ by solving 
    
    \setlength{\parindent}{1em} problem \eqref{optphi.3};
    \STATE \quad Set $\boldsymbol{\xi}_{1,0} = \boldsymbol{\xi}_{1}^{(j)}$ and $\boldsymbol{\xi}_{2,0} = \boldsymbol{\xi}_{2}^{(j)}$;
    \STATE \quad With given $\mathbf{Q}^{(j)}$, $\boldsymbol{\Phi}_d^{(j)}$, and $\boldsymbol{\Phi}_u^{(j)}$ update $\mathbf{p}^{(j)}$ and $\mathbf{g}^{(j)}$ by using \eqref{solution_p.1};
    \STATE \quad With given $\mathbf{Q}^{(j)}$, $\boldsymbol{\Phi}_d^{(j)}$, $\boldsymbol{\Phi}_u^{(j)}$, $\mathbf{p}^{(j)}$, and $\mathbf{g}^{(j)}$, compute $R_{sec}^{(j)}$;
    \STATE \textbf{until}: $\left|R_{sec}^{(j)} - R_{sec}^{(j-1)}\right| \leq \epsilon_c$ or $j>j_{max}$.
    \end{algorithmic} 
\end{algorithm}
Accordingly, problem \eqref{opttraj.1} can be approximately transformed into
\textcolor{black}{\begin{subequations}\label{opttraj.2}
\begin{align}
    & \max\limits_{\mathbf{Q}, \mathbf{u}, \mathbf{e}, \mathbf{s},
     \atop
     \mathbf{t}, \boldsymbol{\zeta}, \mathbf{r}_d, \mathbf{r}_u}  \frac{1}{N}\sum\limits_{n=1}^N\left[w \Hat{R}_{down}^{traj}[n] + \left(1 - w\right) \log_2\left(1+\rho\gamma_1[n]r_u[n]\right)\right] \label{opttraj.2.a}\\
    &\textrm{s.t.} ~ x^2[n]+x^2_G+y^2[n]+y^2_G-2x_G x[n]-2y_G y[n]+z_U^2\nonumber\\
    &\quad~ -\left(1+\frac{4}{\kappa}\right)\left(u_0[n]\right)^{-\frac{4}{\kappa}}+ \frac{4}{\kappa}\left(u_0[n]\right)^{-\frac{4}{\kappa}-1}u[n]\leq0, \forall n, \label{opttraj.2.b}\\
    & x^2[n]+x^2_R+y^2[n]+y^2_R-2x_Rx[n]-\left(1+\frac{4}{\alpha}\right)\left(e_0[n]\right)^{-\frac{4}{\alpha}}\nonumber\\
    & -2y_Ry[n]+(z_U-z_R)^2 + \frac{4}{\alpha}\left(e_0[n]\right)^{-\frac{4}{\alpha}-1}e[n]\leq0, \forall n, \label{opttraj.2.c}\\
    & \left(s[n]\right)^{-\frac{4}{\kappa}}+x_0^2[n] - 2x_0[n]x[n]-x^2_E+y_0^2[n]-2y_0[n]y[n]\nonumber\\
    & -y^2_E+2x_E x[n]+2y_E y[n]-z_U^2 \leq 0, \forall n, \label{opttraj.2.d}\\
    &  \left(t[n]\right)^{-\frac{4}{\alpha}}+x_0^2[n]- 2x_0[n]x[n]-x^2_R+y_0^2[n]-2y_0[n]y[n] \nonumber\\
    & -y^2_R+2x_Rx[n]+2y_Ry[n]-(z_U-z_R)^2 \leq 0, \forall n, \label{opttraj.2.e}\\
    & r_d[n]+\Tilde{\boldsymbol{h}}_{ue,0}^T[n]\mathbf{H}_{QG}[n]\Tilde{\boldsymbol{h}}_{ue,0}[n] \nonumber\\ 
    &\qquad~ -2\Re\left[\Tilde{\boldsymbol{h}}_{ue,0}^T[n]\mathbf{H}_{QG}[n]\Tilde{\boldsymbol{h}}_{ue}[n]\right]\leq 0, \forall n, \label{opttraj.2.f}\\
    & r_u[n]+\Tilde{\boldsymbol{h}}_{ue,0}^T[n]\mathbf{H}_{GQ}[n]\Tilde{\boldsymbol{h}}_{ue,0}[n]\nonumber\\
    &\qquad~- 2\Re\left[\Tilde{\boldsymbol{h}}_{ue,0}^T[n]\mathbf{H}_{GQ}[n]\Tilde{\boldsymbol{h}}_{ue}[n]\right]\leq 0, \forall n, \label{opttraj.2.g}\\
    & \eqref{mobility.1}, \eqref{opttraj.1.f}, \eqref{opttraj.1.i}, \nonumber
\end{align}
\end{subequations}}
where
$$\Hat{R}_{down}^{traj}[n] = \log_2\left(1+\rho\gamma_0[n]r_d[n]\right) - \frac{\zeta[n]}{\ln{2}\left(1 + \zeta_0[n] \right)}.$$
Problem \eqref{opttraj.2} is a convex optimization problem, and thus the CVX solver can be used to obtain the solution.

\subsection{Overall Algorithm}
With the proposed solutions to the three sub-problems, the overall algorithm for solving problem \eqref{optimal.1} is summarized in Algorithm \ref{alg1}, \textcolor{black}{where $\epsilon_c$ is used to control the accuracy of convergence, and $j_{max}$ denotes as the maximum number of iterations.} Solving sub-problem 2 and sub-problem 3 by using the interior-point method dominates the complexity of Algorithm \ref{alg1}. Based on the results in \cite{yu2019robust} and \cite{you2020}, the computational complexities of solving sub-problem 2 and sub-problem 3 are
\begin{equation}
    \begin{split}
        \mathcal{O}_{sub2}\left(2\sqrt{M+1}\log\left(1/\epsilon_c\right)\left(2N\left(M+1\right)^3 \right.\right.\\
        \left.\left.+ 4N^2\left(M+1\right)^2 + 8N^3\right)\right)\nonumber
    \end{split}
\end{equation}
and $\mathcal{O}_{sub3}\left(\left(8N\right)^{3.5}\log\left(1/\epsilon_c\right)\right)$, respectively. Hence, the overall complexity of solving problem \eqref{optimal.1} is $\mathcal{O}_{sub2} + \mathcal{O}_{sub3}$. Furthermore, as shown in Fig.~\ref{fig_c}, we observe that the proposed algorithm can quickly converge.

\section{Simulation Results}
In this section, we present simulation results to verify the validity of the proposed algorithm (denoted as JO) for the joint UL/DL optimization. The following benchmark algorithms are used for comparison:
\begin{itemize}
    \item Robust design of the UAV trajectory and transmit power without passive beamforming (denoted as JO/NPB).
    \item Robust design of the heuristic trajectory, transmit power, and passive beamforming (denoted as JO/HT).
    \item Non-robust design of the UAV trajectory, passive beamforming, and transmit power (denoted as JO/NR).
\end{itemize}
Specifically, ``heuristic trajectory'' refers to a preset trajectory where the UAV first flies directly to the ground user at the maximum speed, then hovers above the user as long as possible, and finally flies to the final location at its maximum speed for the rest of the flight time. Also, for the considered JO/NR algorithm, the estimated CSI of the eavesdropping channels is the exact CSI. Hence, it is a special case of our proposed algorithm that is obtained by setting $\epsilon_1 = \epsilon_2 = 0$. From the definitions of the uncertainty radii $\epsilon_1$ and $\epsilon_2$ in \cite{yu2019robust}, the maximum normalized estimation error of the eavesdropping links is defined as $\delta_l = \epsilon_l / \lVert \Bar{\boldsymbol{h}}_{El}\rVert$, where $l \in \{1, 2\}$. Since the UAV usually flies higher than the RIS, the ground user and the eavesdropper in the DL transmission, the Rician factors for the U-G and U-E links are set to $\beta_{U\!G} = \beta_{U\!E} = 10$ $\rm{dB}$, while the Rician factors for the R-G, R-E, and U-R links are set to $\beta_{U\!R} = \beta_{RG} = \beta_{RE} = 3$ $\rm{dB}$. The corresponding Rician factors in the UL transmission are similar to the DL transmission, i.e., $\beta_{GU} = 10$ $\rm{dB}$ and $\beta_{RU} = \beta_{G\!R} = \beta_{GE} = 3$ $\rm{dB}$. The initial feasible solutions of our proposed JO algorithm is given by the JO/HT algorithm. The remaining parameters are as follows: $\mathbf{q}_0 = [-500, 20]^T$ $\rm{m}$, $\mathbf{q}_F = [500, 20]^T$ $\rm{m}$, $\mathbf{w}_G=[0, 120]^T$ $\rm{m}$, $\mathbf{w}_E=[200, 150]^T$ $\rm{m}$, $\mathbf{w}_R=[0, 0]^T$ $\rm{m}$, $z_U=100$ $\rm{m}$, $z_R=40$ $\rm{m}$, $v_{max}=30$ $\rm{m/s}$, \textcolor{black}{$\delta_t=0.4$ $\rm{s}$,} $M=M_x \times M_y = 6 \times 5$, $\sigma^2=-80$ $\rm{dBm}$, $d=\frac{\lambda}{2}$, $\alpha=2.2$, $\kappa=3.3$, $\varsigma = 3.4$, $\rho=-30$ $\rm{dB}$, $\epsilon_c = 10^{-3}$, $j_{max} = 40$, $P_{peak} = 4\Bar{P}$, and $G_{peak} = 4\Bar{G}$. In paticular, we assume that all wiretap channels have the same maximum normalized estimation error variance, namely, $\delta_1 = \delta_2 = \delta_a$.

\begin{figure}
    \centering
    \includegraphics[width=3.0in]{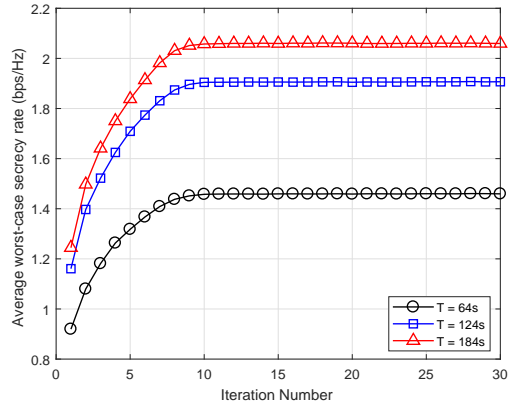}
    \caption{Average worst-case secrecy rate versus iteration number}
    \label{fig_c}
\end{figure}

Fig.~\ref{fig_c} plots the average worst-case secrecy rate of the proposed algorithm versus the number of iterations under different flight periods with $w = 0.5$, $\delta_a^2=0.5$ and $\Bar{P}=\Bar{G}=20$ $\rm{dBm}$. It is observed that our proposed algorithm can quickly converge after around 10 iterations, and the average worst-case secrecy rate increases by increasing the flight time $T$.

\begin{figure*}[htp]
    \centering
    \subfigure[The UAV trajectory of the JO algorithm]{\label{fig2:a}
    \includegraphics[width=3.0in]{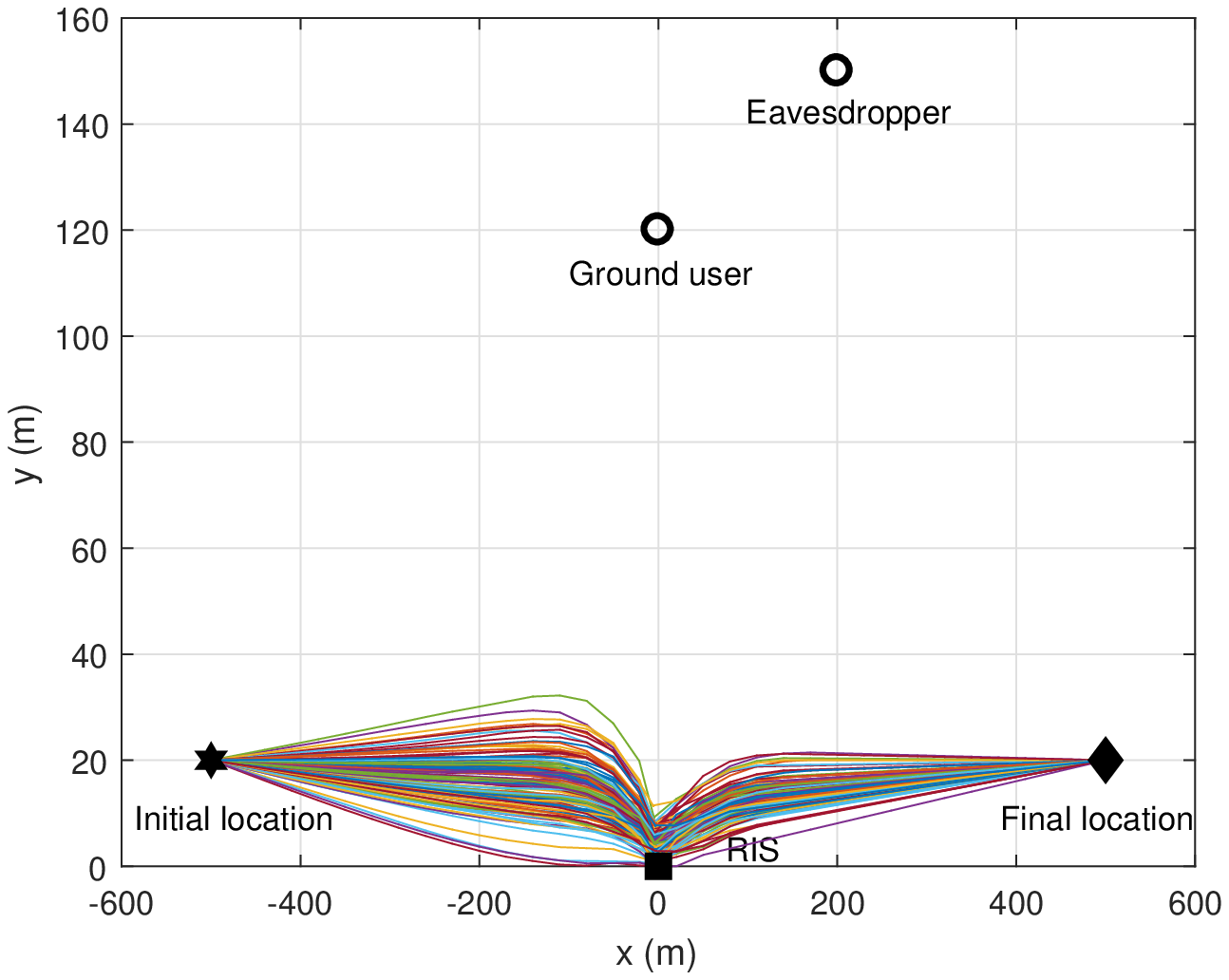}}
    \subfigure[The UAV trajectory of the JO/NPB algorithm]{
    \label{fig2:b}
    \includegraphics[width=3.0in]{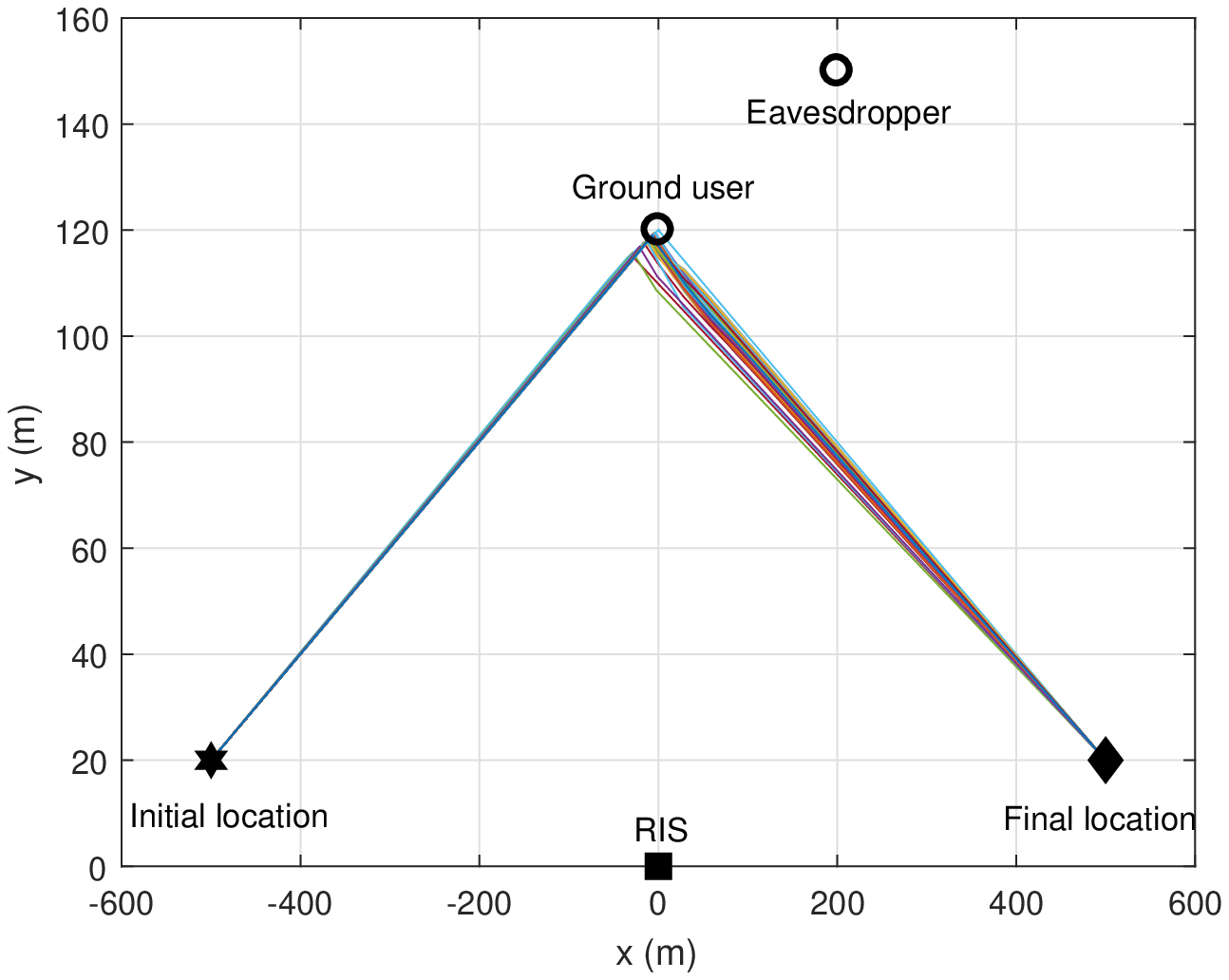}
    }
    \quad
    \subfigure[The UAV trajectory of the JO/NR algorithm]{
    \label{fig2:c}
    \includegraphics[width=3.0in]{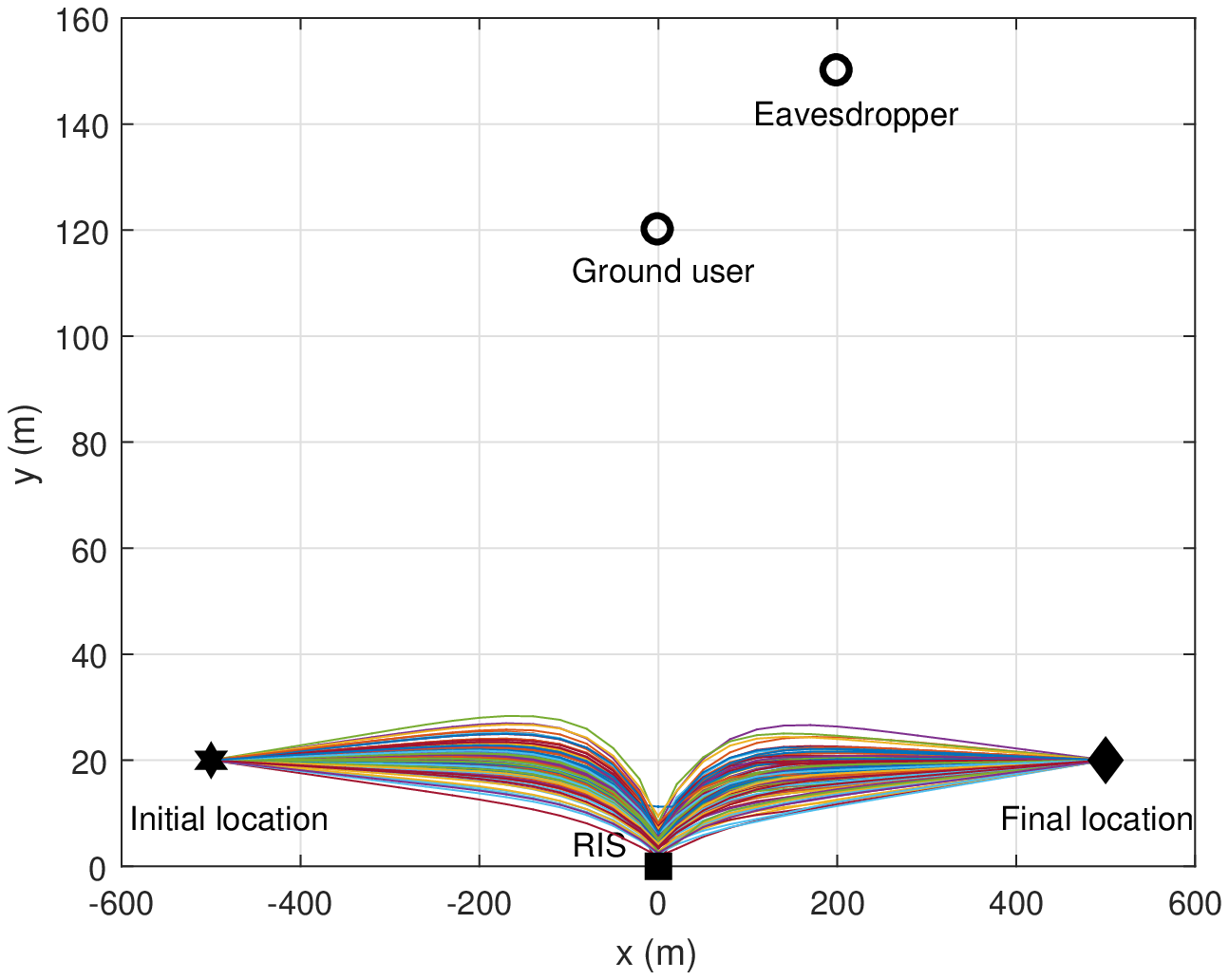}
    }
    \caption{UAV trajectories by using different algorithms with $T = 124$ $\rm{s}$, $\delta_a^2=0.5$, $\Bar{P}=\Bar{G}=20$ $\rm{dBm}$, and $w = 0.5$}
    \label{fig2}
\end{figure*}

\begin{figure}
    \centering
    \includegraphics[width=3.0in]{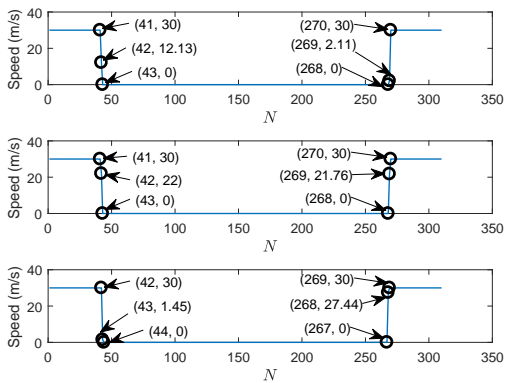}
    \caption{\textcolor{black}{Upper figure: UAV speed (m/s) versus $N$ for the JO algorithm; middle figure: UAV speed (m/s) versus $N$ for the JO/NR algorithm; lower figure: UAV speed (m/s) versus $N$ for JO/NPB algorithm. The system parameters are set as $T = 124$ $\rm{s}$, $\delta_a^2=0.5$, $\Bar{P}=\Bar{G}=20$ $\rm{dBm}$, and $w = 0.5$.}}
    \label{fig_s}
\end{figure}

In Fig.~\ref{fig2}, we illustrate the UAV trajectories for different algorithms by setting $T = 124$ $\rm{s}$, $\delta_a^2=0.5$, $\Bar{P}=\Bar{G}=20$ $\rm{dBm}$, and $w=0.5$. Fig.~\ref{fig2:a}, Fig.~\ref{fig2:b}, and Fig.~\ref{fig2:c} show more than 100 UAV trajectories\footnote{Since all channels are modeled by the Rician fading channel model, the optimized UAV trajectories in different random independent realizations of all channels are different. Hence, for different algorithms, we draw more than 100 UAV trajectories to show the general trajectory trends of them.} by using the JO, JO/NPB and JO/NR algorithms, respectively, when $T$ is sufficiently large (e.g., $T = 124$ $\rm{s}$). \textcolor{black}{In Fig.~\ref{fig_s}, we show the UAV speed during a period of $T = 124$ $\rm{s}$ for different algorithms}\footnote{\textcolor{black}{Unlike Fig.~\ref{fig2}, for different algorithms, we only show the UAV speed versus $N$ for different algorithms under a certain random realization. This is because for each algorithm, the curves of the UAV speed remain almost the same under different random realizations.}}. \textcolor{black}{From Fig.~\ref{fig2} and Fig.~\ref{fig_s}, it is observed that, for all the algorithms, the UAV first flies to a certain location at $V_{max}$, then keeps static, and finally flies to the final location at $V_{max}$. However,} as for the JO/NPB algorithm, the UAV first flies directly to a certain location where the UAV is close to the ground user and away from the eavesdropper as possible as it can, then hovers as long as possible, and finally flies along a relatively direct path to the final location in order to avoid being eavesdropped. By contrast, for the JO and JO/NR algorithms, the UAV tends to fly along an arc path to a certain location between the ground user and the RIS, then it hovers as long as possible, and finally reaches the final location. This is because the JO and JO/NR algorithms balance the channel gains between the direct links (i.e., the U-G, U-E, G-U, and G-E links) and reflecting links (i.e., the U-R-G, U-R-E, G-R-U, and G-R-E links) in each time slot in order to choose a trajectory, so as to achieve the best communication quality. Besides, since the JO algorithm takes the CSI uncertainty into account, the UAV trajectories of the JO algorithm under different channel realizations are more decentralized than those of the JO/NR algorithm under different channel realizations.


\begin{figure}
    \centering
    \includegraphics[width=3.0in]{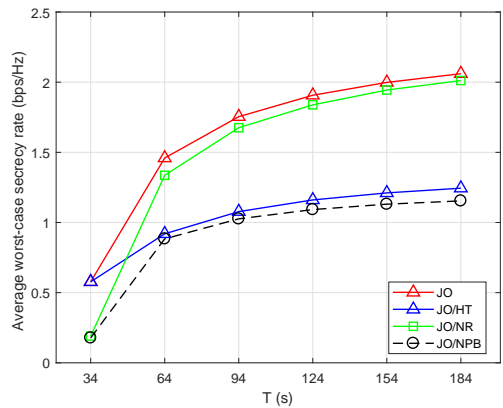}
    \caption{Average worst-case sccrecy rate performance by different algorithms versus $T$}
    \label{fig3}
\end{figure}

In Fig.~\ref{fig3}, we show the average worst-case secrecy rates for different algorithms versus $T$ with $w=0.5$, $\delta_a^2=0.5$, and $\Bar{P}=\Bar{G}=20$ $\rm{dBm}$. At the hovering location, the trade-off between enhancing the quality of the legitimate links and weakening the quality of the eavesdropping links is achieved for the UAV. Therefore, the maximum secrecy rate is achieved at the hovering location and the longer the UAV remains static at the hovering location, the larger the average worst-case secrecy rate is. This is the reason why the average worst-case secrecy rates of all the algorithms increase with $T$. In particular, our proposed algorithm exceeds all the benchmark schemes. This shows that, with the aid of the proposed robust joint design of the UAV trajectory, RIS's passive beamforming, and transmit power control, the secrecy rate performance can be effectively improved. Furthermore, it is worth noting that the JO/NR algorithm outperforms the other benchmark algorithms. This demonstrates that even though the CSI uncertainty of the eavesdropping channels is not taken into account, which leads to inaccurate optimization, the joint design of the UAV trajectory, passive beamforming, and transmit power can still achieve a considerable gain, as compared with the counterpart schemes. 

\begin{figure}
    \centering
    \includegraphics[width=3.0in]{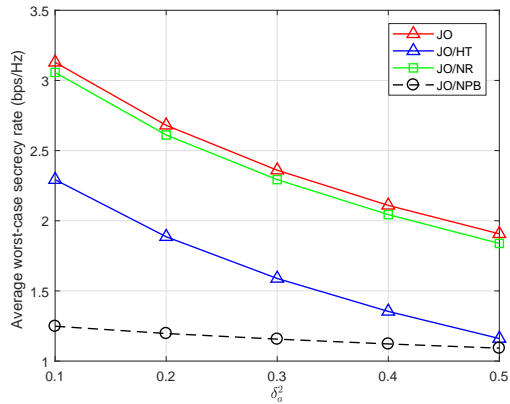}
    \caption{Average worst-case secrecy rate versus the maximum normalized channel estimation error variance}
    \label{fig4}
\end{figure}
In Fig.~\ref{fig4}, we investigate the average worst-case secrecy rates for different algorithms versus $\delta_a^2$ with $w=0.5$, $T = 124$ $\rm{s}$, and $\Bar{P}=\Bar{G}=20$ $\rm{dBm}$. We observe that the average worst-case secrecy rates of all the algorithms decrease as the CSI uncertainty of the wiretap channels increases. This is because large values of the CSI uncertainty of the wiretap channels make it more difficult to achieve a robust design. \textcolor{black}{However, capitalizing on the proposed robust joint design, the JO algorithm achieves a better secrecy rate performance than the other benchmark algorithms.} Furthermore, the secrecy rate performance of the JO algorithm exceeds that of the JO/NR algorithm, which demonstrates that our proposed algorithm is robust. In addition, it is observed that although the CSI estimation errors of the eavesdropping links are not taken into account, the average worst-case secrecy rate of the JO/NR algorithm still exceeds that of the other benchmark schemes. Once again, this demonstrates that the joint design of the UAV trajectory, passive beamforming, and transmit power achieves a substantial gain. Besides, it is worth noting that the secrecy rate performance of the JO/NPB algorithm is close to that of the JO/HT algorithm when $\delta_a^2$ is sufficiently large (e.g., $\delta_a^2 = 0.5$). This is mainly because too large CSI uncertainty leads to the failure of RIS's passive beamforming, even to the reverse effect. By contrast, large CSI uncertainty has a marginal effect on the trajectory or transmit power optimization, which is demonstrated by the smooth curve of the JO/NPB algorithm in Fig.~\ref{fig4}.

\begin{figure*}[htp]
    \centering
    \subfigure[$w = 0.1$]{\label{fig5:a}
    \includegraphics[width=3.0in]{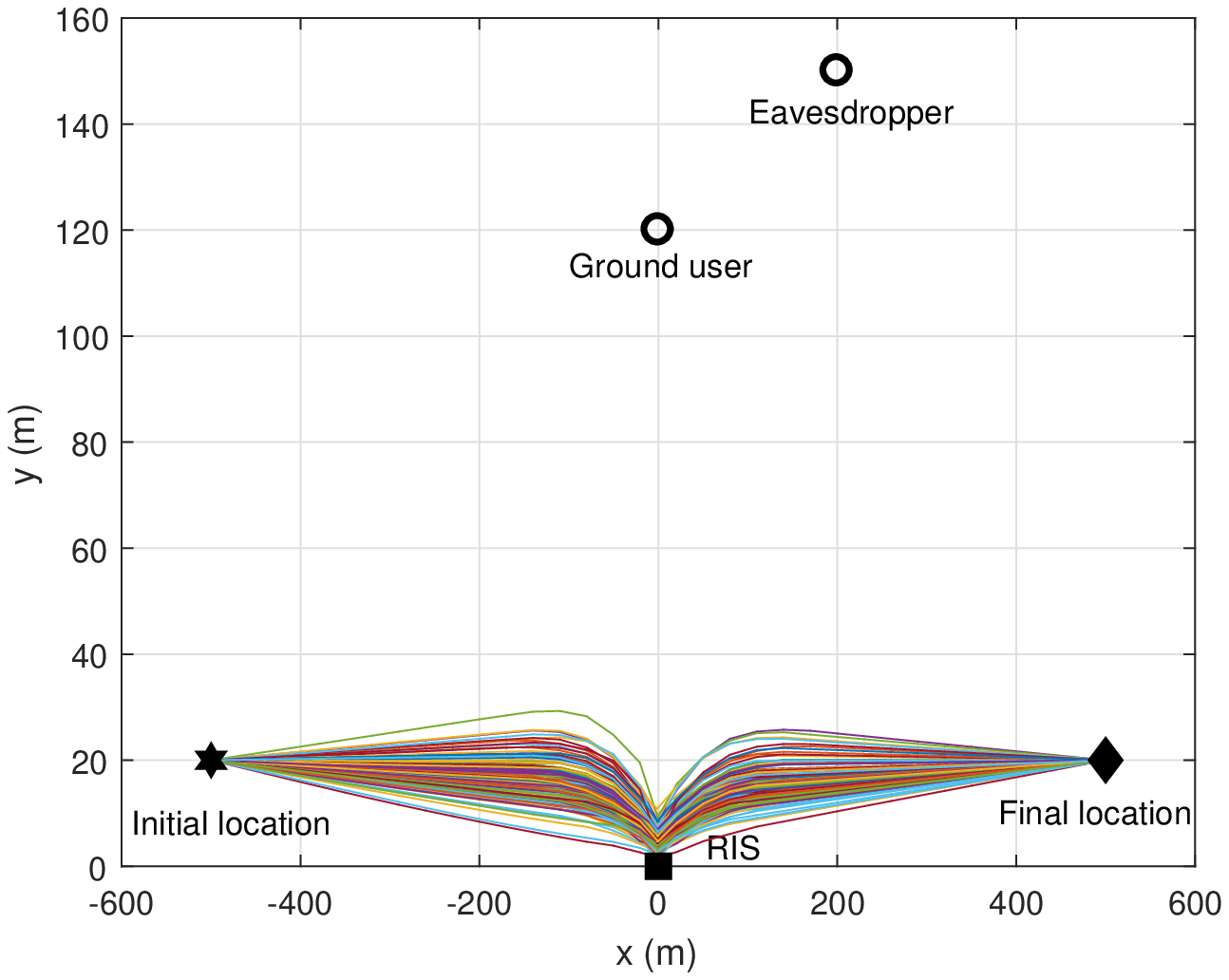}}
    \subfigure[$w = 0.3$]{
    \label{fig5:b}
    \includegraphics[width=3.0in]{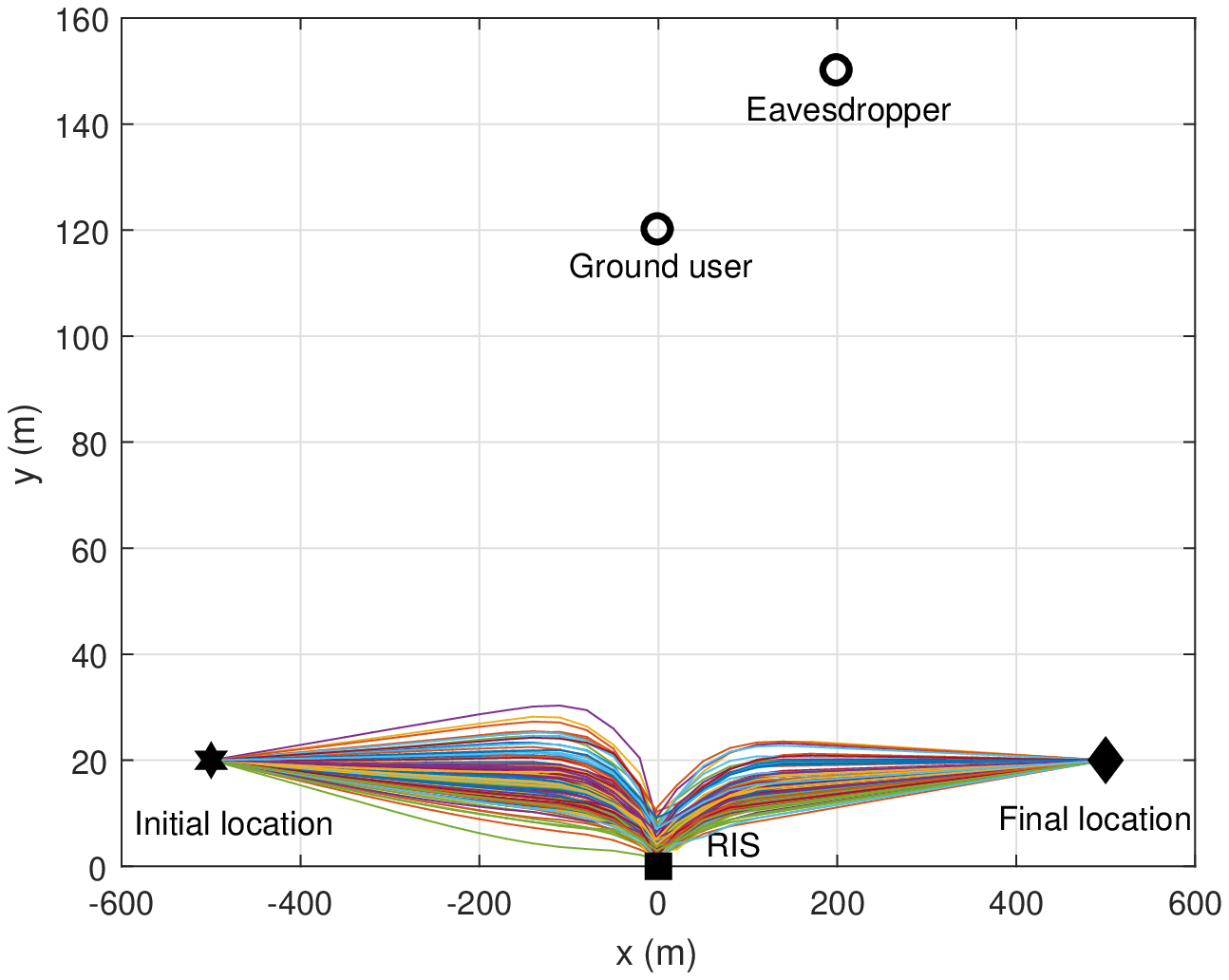}
    }
    \quad
    \subfigure[$w = 0.7$]{
    \label{fig5:c}
    \includegraphics[width=3.0in]{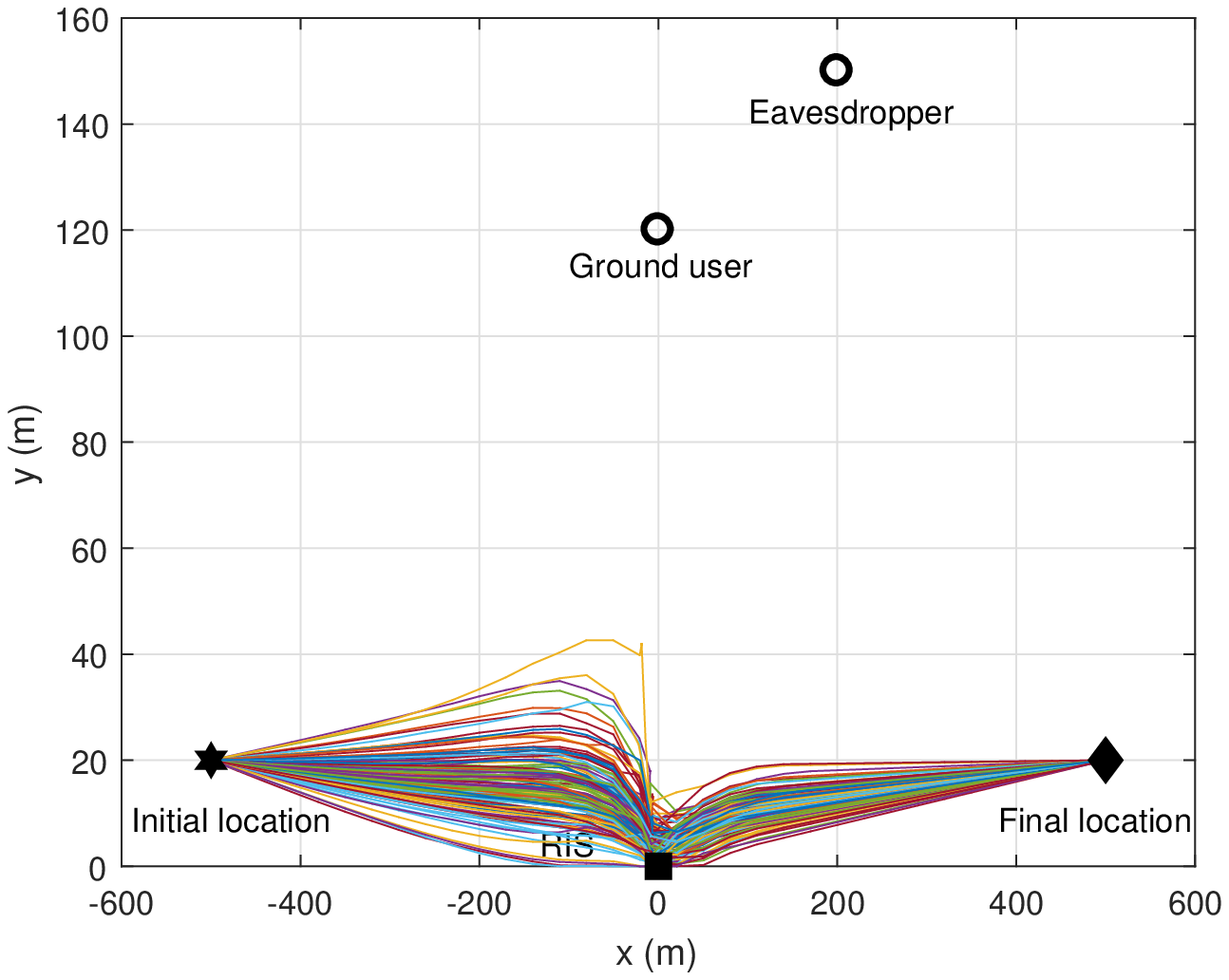}
    }
    \subfigure[$w = 0.9$]{
    \label{fig5:d}
    \includegraphics[width=3.0in]{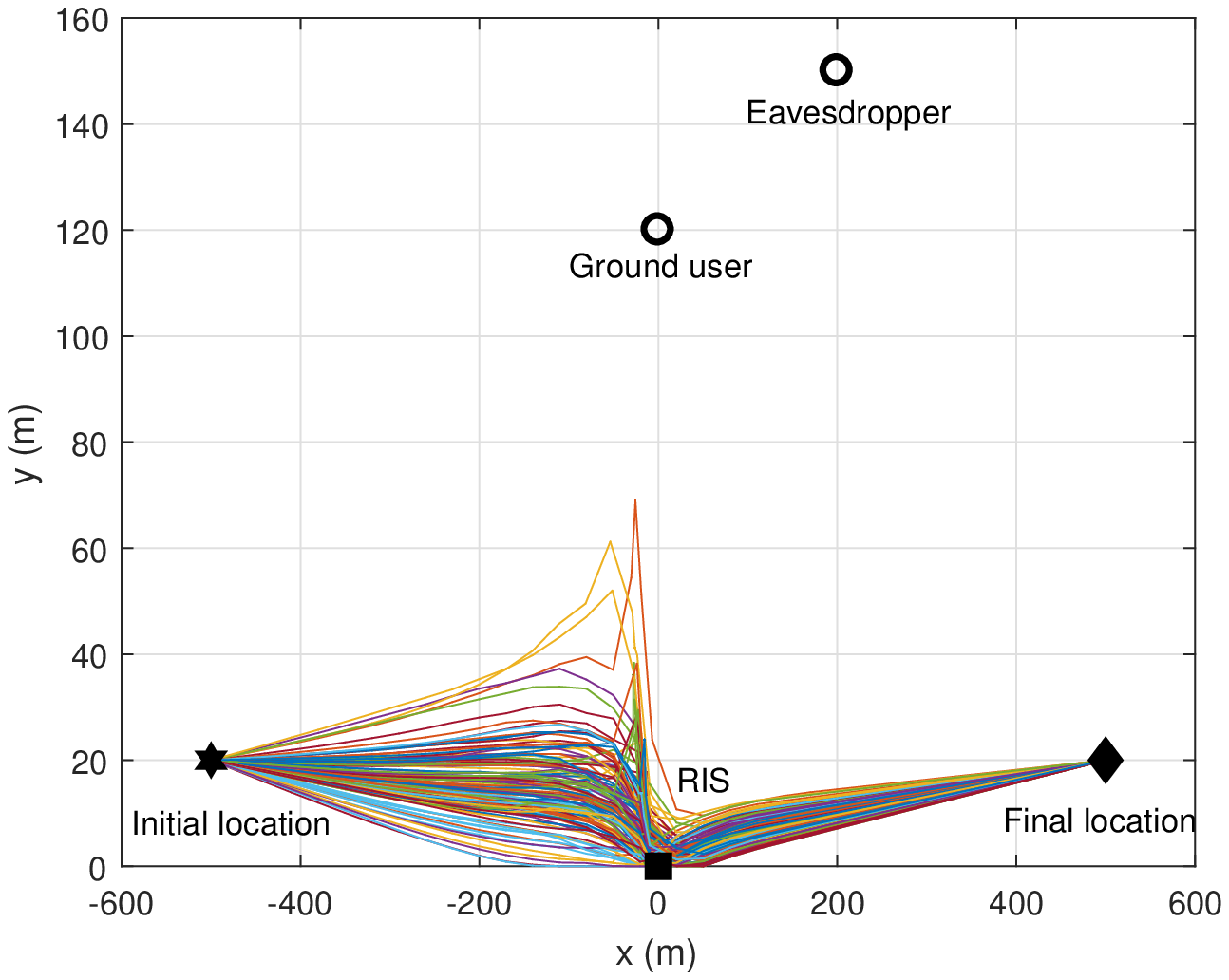}
    }
    \caption{UAV trajectories of the JO algorithm by different $w$ with $T = 124$ $\rm{s}$, $\delta_a^2=0.5$, $\Bar{P}=\Bar{G}=20$ $\rm{dBm}$}
    \label{fig5}
\end{figure*}

In Fig.~\ref{fig5}, we show the UAV trajectories by different time slot division setups, $w$, with $T = 124$ $\rm{s}$, $\delta_a^2=0.5$, and $\Bar{P}=\Bar{G}=20$ $\rm{dBm}$. In particular, $w = 0.1$ means that we pay more attention to the UL communications, while $w = 0.9$ means that we focus more on the DL communications. Since $R_{G\!E}[n]$ is independent of the UAV trajectory, in the UL communications, the UAV trajectory is only designed for the maximum of the achievable rate $R_{GU}[n]$. Hence, for $w = 0.1$, the UL communication is dominant, and the JO algorithm almost achieves the trade-off between the channel gains of the G-U link and G-R-U link to choose a trajectory, so as to achieve the best communication quality. When $w$ increases, the DL communication becomes more and more dominant, it is more important for the JO algorithm to balance the channel gains between the U-G link and U-R-G link and between the U-E link and U-R-E link to design the UAV trajectory. Hence, the JO algorithm not only considers how to increase the legitimate rates between the UAV and the ground user, but also considers how to decrease the wiretap rate from the UAV to the eavesdropper. This is also the reason why the first half paths, i.e., the paths from the initial location to the hovering location, when $w=0.9$ are more decentralized than those when $w=0.1$. Besides, in the second half paths, i.e., the paths from the hovering location to the final location, the UAV is closer to the eavesdropper than in the first half paths. Thus, when $w$ is sufficiently large (e.g., $w=0.9$), the UAV is inclined to fly along relatively direct paths to the final location, so as to avoid the information leakage and increase the secrecy rate.

\section{Conclusion}
In this paper, \textcolor{black}{as a supplement of the higher-layer encryption techniques, we studied a novel RIS-assisted UAV physical-layer secure communication}, aiming at integrating RIS and UAV technologies for improving the system secrecy rate. In particular, a single flight time slot is allocated to the DL and UL transmissions between the UAV and the ground user, while the legitimate channels are wiretapped by an eavesdropper. Since the eavesdroppers always avoid being detected by the legitimate transmitter, the acquisition of the CSI of the eavesdropping channels is usually imperfect. Thus, we focused our attention on the joint and CSI-robust design of the UAV's trajectory, RIS's passive beamforming, and transmit power of the legitimate transmitter in order to maximize the average worst-case secrecy rate of the considered communication system. Although the formulated problem is intractable due to its non-convexity, we proposed an efficient algorithm to approximately solve it by applying the AO, SCA, $\mathcal{S}$-Procedure, and SDR techniques. Simulation results demonstrated that the assistance of an RIS is beneficial to substantially improve the secrecy rate performance, and the joint design of UAV trajectory, RIS's passive beamforming, transmit power of the legitimates can achieve a substantial gain. In addition, the robustness of our proposed algorithm was confirmed with respect to inaccurate estimates of the CSI of the wiretap channels.

\textcolor{black}{The findings of this paper can be used as a reference for the study of the multi-user scenario. We now briefly discuss the extension of our proposed scheme to the multi-user scenario. Clearly, if an orthogonal multiple access protocol is employed, our proposed single-user scheme can be directly applied, except that an effective user scheduling strategy should be developed to schedule the data transmission of multiple users. If a non-orthogonal multiple access protocol is considered, then our proposed scheme cannot be directly applied and an effective interference cancellation algorithm is required to suppress the interference between multiple users. However, a detailed discussion of this issue is beyond the scope of this paper, and we leave this as future work.}

\bibliographystyle{ieeetr}
\bibliography{myreference}

\begin{thebibliography}{10}

\bibitem{macro2019}
E.~{Basar}, M.~{Di Renzo}, J.~{De Rosny}, M.~{Debbah}, M.~{Alouini}, and
  R.~{Zhang}, ``Wireless communications through reconfigurable intelligent
  surfaces,'' {\em IEEE Access}, vol.~7, pp.~116753--116773, 2019.

\bibitem{Gupta2016}
L.~{Gupta}, R.~{Jain}, and G.~{Vaszkun}, ``Survey of important issues in {UAV}
  communication networks,'' {\em IEEE Commun. Surveys Tuts.}, vol.~18, no.~2,
  pp.~1123--1152, 2016.

\bibitem{zeng2017}
Y.~{Zeng} and R.~{Zhang}, ``Energy-efficient {UAV} communication with
  trajectory optimization,'' {\em IEEE Trans. Wireless Commun.}, vol.~16,
  no.~6, pp.~3747--3760, 2017.

\bibitem{zhang2019}
G.~{Zhang}, Q.~{Wu}, M.~{Cui}, and R.~{Zhang}, ``Securing {UAV} communications
  via joint trajectory and power control,'' {\em IEEE Trans. Wireless Commun.},
  vol.~18, pp.~1376--1389, Feb. 2019.

\bibitem{macro42020}
X.~{Qian}, M.~{Di Renzo}, J.~{Liu}, A.~{Kammoun}, and M.~{Alouini},
  ``Beamforming through reconfigurable intelligent surfaces in single-user
  {MIMO} systems: {SNR} distribution and scaling laws in the presence of
  channel fading and phase noise,'' {\em \rm {[Online] Available:
  https://arxiv.org/abs/2005.07472v1.}}

\bibitem{yan22020}
W.~{Yan}, X.~{Yuan}, and X.~{Kuai}, ``Passive beamforming and information
  transfer via large intelligent surface,'' {\em IEEE Wireless. Commun. Lett.},
  vol.~9, no.~4, pp.~533--537, 2020.

\bibitem{yan32020}
W.~{Yan}, X.~{Yuan}, Z.~{He}, and X.~{Kuai}, ``Large intelligent surface aided
  multiuser {MIMO}: Passive beamforming and information transfer,'' in {\em ICC
  2020 - 2020 IEEE International Conference on Communications (ICC)}, pp.~1--7,
  2020.

\bibitem{macro72020}
A.~Zappone, M.~{Di Renzo}, F.~Shams, X.~Qian, and M.~Debbah, ``Overhead-aware
  design of reconfigurable intelligent surfaces in smart radio environments,''
  {\em \rm {[Online] Available: https://arxiv.org/abs/2003.02538.}}

\bibitem{macro52020}
G.~{Zhou}, C.~{Pan}, H.~{Ren}, K.~{Wang}, M.~{Di Renzo}, and A.~{Nallanathan},
  ``Robust beamforming design for intelligent reflecting surface aided {MISO}
  communication systems,'' {\em IEEE Wireless. Commun. Lett.}, pp.~1--1, 2020.

\bibitem{Schober2020}
D.~{Xu}, Y.~{Sun}, D.~W.~K. {Ng}, and R.~{Schober}, ``Multiuser {MISO} {UAV}
  communications in uncertain environments with no-fly zones: Robust trajectory
  and resource allocation design,'' {\em IEEE Trans. Commun.}, vol.~68, no.~5,
  pp.~3153--3172, 2020.

\bibitem{zeng2019}
Y.~{Zeng}, Q.~{Wu}, and R.~{Zhang}, ``Accessing from the sky: A tutorial on
  {UAV} communications for {5G} and beyond,'' {\em Proc. IEEE}, vol.~107,
  no.~12, pp.~2327--2375, 2019.

\bibitem{zhu2020}
L.~Zhu, J.~Zhang, Z.~Xiao, X.~Cao, X.-G. Xia, and R.~Schober, ``Millimeter-wave
  full-duplex {UAV} relay: Joint positioning, beamforming, and power control,''
  {\em \rm {[Online] Available: https://arxiv.org/abs/2004.11070.}}

\bibitem{lian2019}
A.~{Li}, Q.~{Wu}, and R.~{Zhang}, ``{UAV}-enabled cooperative jamming for
  improving secrecy of ground wiretap channel,'' {\em IEEE Wireless. Commun.
  Lett.}, vol.~8, no.~1, pp.~181--184, 2019.

\bibitem{zhong2019}
C.~{Zhong}, J.~{Yao}, and J.~{Xu}, ``Secure {UAV} communication with
  cooperative jamming and trajectory control,'' {\em IEEE Wireless. Commun.
  Lett.}, vol.~23, no.~2, pp.~286--289, 2019.

\bibitem{lee2018}
H.~{Lee}, S.~{Eom}, J.~{Park}, and I.~{Lee}, ``{UAV}-aided secure
  communications with cooperative jamming,'' {\em IEEE Trans. Veh. Technol.},
  vol.~67, no.~10, pp.~9385--9392, 2018.

\bibitem{cai2018}
Y.~{Cai}, F.~{Cui}, Q.~{Shi}, M.~{Zhao}, and G.~Y. {Li}, ``Dual-{UAV} enabled
  secure communications: Joint trajectory design and user scheduling,'' {\em
  IEEE J. Sel. Areas Commun.}, vol.~36, no.~9, pp.~1972--1985, 2018.

\bibitem{zhou2019}
X.~{Zhou}, Q.~{Wu}, S.~{Yan}, F.~{Shu}, and J.~{Li}, ``{UAV}-enabled secure
  communications: Joint trajectory and transmit power optimization,'' {\em IEEE
  Trans. Veh. Technol.}, vol.~68, no.~4, pp.~4069--4073, 2019.

\bibitem{cui22018}
M.~{Cui}, G.~{Zhang}, Q.~{Wu}, and D.~W.~K. {Ng}, ``Robust trajectory and
  transmit power design for secure {UAV} communications,'' {\em IEEE Trans.
  Veh. Technol.}, vol.~67, no.~9, pp.~9042--9046, 2018.

\bibitem{macro2020smart}
M.~{Di Renzo} {\em et~al.}, ``Smart radio environments empowered by
  reconfigurable intelligent surfaces: How it works, state of research, and
  road ahead,'' {\em \rm {[Online] Available:
  https://arxiv.org/abs/2004.09352.}}

\bibitem{macro2019smart}
M.~{Di Renzo} {\em et~al.}, ``Smart radio environments empowered by
  reconfigurable {AI} meta-surfaces: An idea whose time has come,'' {\em
  EURASIP J. Wireless Commun. Net.}, vol.~2019, p.~129, May 2019.

\bibitem{macro62020}
C.~{Huang}, S.~{Hu}, G.~C. {Alexandropoulos}, A.~{Zappone}, C.~{Yuen},
  R.~{Zhang}, M.~{Di Renzo}, and M.~{Debbah}, ``Holographic {MIMO} surfaces for
  6{G} wireless networks: Opportunities, challenges, and trends,'' {\em IEEE
  Wirel. Commun. Mag.}, pp.~1--8, 2020.

\bibitem{yuan2020}
X.~Yuan, Y.-J. Zhang, Y.~Shi, W.~Yan, and H.~Liu,
  ``Reconfigurable-intelligent-surface empowered 6{G} wireless communications:
  Challenges and opportunities,'' {\em \rm {[Online] Available:
  https://arxiv.org/abs/2001.00364.}}

\bibitem{wu2019intelligent}
Q.~{Wu} and R.~{Zhang}, ``Intelligent reflecting surface enhanced wireless
  network via joint active and passive beamforming,'' {\em IEEE Trans. Wireless
  Commun.}, vol.~18, pp.~5394--5409, Nov. 2019.

\bibitem{guan2020}
X.~{Guan}, Q.~{Wu}, and R.~{Zhang}, ``Intelligent reflecting surface assisted
  secrecy communication: Is artificial noise helpful or not?,'' {\em IEEE
  Wireless. Commun. Lett.}, vol.~9, no.~6, pp.~778--782, 2020.

\bibitem{yu2019globe}
X.~{Yu}, D.~{Xu}, and R.~{Schober}, ``Enabling secure wireless communications
  via intelligent reflecting surfaces,'' in {\em 2019 IEEE Global
  Communications Conference (GLOBECOM)}, pp.~1--6, 2019.

\bibitem{xu2020}
D.~{Xu}, X.~{Yu}, Y.~{Sun}, D.~W.~K. {Ng}, and R.~{Schober}, ``Resource
  allocation for secure {IRS}-assisted multiuser {MISO} systems,'' in {\em 2019
  IEEE Globecom Workshops (GC Wkshps)}, pp.~1--6, 2019.

\bibitem{lu2020}
X.~{Lu}, W.~{Yang}, X.~{Guan}, Q.~{Wu}, and Y.~{Cai}, ``Robust and secure
  beamforming for intelligent reflecting surface aided mmwave {MISO} systems,''
  {\em IEEE Wireless. Commun. Lett.}, pp.~1--1, 2020.

\bibitem{yu2019robust}
X.~{Yu}, D.~{Xu}, Y.~{Sun}, D.~W.~K. {Ng}, and R.~{Schober}, ``Robust and
  secure wireless communications via intelligent reflecting surfaces,'' {\em
  IEEE J. Sel. Areas Commun.}, vol.~38, no.~11, pp.~2637--2652, 2020.

\bibitem{macro32020}
L.~{Yang}, F.~{Meng}, J.~{Zhang}, M.~O. {Hasna}, and M.~{Di Renzo}, ``On the
  performance of {RIS}-assisted dual-hop {UAV} communication systems,'' {\em
  IEEE Trans. Veh. Technol.}, pp.~1--1, 2020.

\bibitem{li2020}
S.~{Li}, B.~{Duo}, X.~{Yuan}, Y.-C. {Liang}, and M.~{Di Renzo},
  ``Reconfigurable intelligent surface assisted {UAV} communication: Joint
  trajectory design and passive beamforming,'' {\em IEEE Wireless. Commun.
  Lett.}, vol.~9, no.~5, pp.~716--720, 2020.

\bibitem{qian2019}
Q.~{Zhang}, W.~{Saad}, and M.~{Bennis}, ``Reflections in the sky: Millimeter
  wave communication with {UAV}-carried intelligent reflectors,'' in {\em 2019
  IEEE Global Communications Conference (GLOBECOM)}, pp.~1--6, 2019.

\bibitem{long2020}
H.~Long, M.~Chen, Z.~Yang, B.~Wang, Z.~Li, X.~Yun, and M.~Shikh-Bahaei,
  ``Reflections in the sky: Joint trajectory and passive beamforming design for
  secure {UAV} networks with reconfigurable intelligent surface,'' {\em \rm
  {[Online] Available: https://arxiv.org/abs/2005.10559}}.

\bibitem{wang2020joint}
L.~Wang, K.~Wang, C.~Pan, W.~Xu, and N.~Aslam, ``Joint trajectory and passive
  beamforming design for intelligent reflecting surface-aided {UAV}
  communications: A deep reinforcement learning approach,'' {\em \rm {[Online]
  Available: https://arxiv.org/abs/2007.08380}}.

\bibitem{ge2020}
L.~{Ge}, P.~{Dong}, H.~{Zhang}, J.~{Wang}, and X.~{You}, ``Joint beamforming
  and trajectory optimization for intelligent reflecting surfaces-assisted
  {UAV} communications,'' {\em IEEE Access}, vol.~8, pp.~78702--78712, 2020.

\bibitem{hua2020}
M.~Hua, L.~Yang, Q.~Wu, C.~Pan, C.~Li, and A.~L. Swindlehurst, ``{UAV}-assisted
  intelligent reflecting surface symbiotic radio system,'' {\em \rm {[Online]
  Available: https://arxiv.org/abs/2007.14029}}.

\bibitem{access1}
M.~B. {Shahab}, R.~{Abbas}, M.~{Shirvanimoghaddam}, and S.~J. {Johnson},
  ``Grant-free non-orthogonal multiple access for {IoT}: A survey,'' {\em IEEE
  Commun. Surveys Tuts.}, vol.~22, no.~3, pp.~1805--1838, 2020.

\bibitem{vision1}
S.~{Minaeian}, J.~{Liu}, and Y.~{Son}, ``Vision-based target detection and
  localization via a team of cooperative {UAV} and {UGVs},'' {\em IEEE Trans.
  Syst., Man, Cybern., Syst.}, vol.~46, no.~7, pp.~1005--1016, 2016.

\bibitem{vision2}
S.~{Sohn}, B.~{Lee}, J.~{Kim}, and C.~{Kee}, ``Vision-based real-time target
  localization for single-antenna {GPS}-guided {UAV},'' {\em IEEE Trans. Aero.
  Elec. Sys.}, vol.~44, no.~4, pp.~1391--1401, 2008.

\bibitem{schober2014}
D.~W.~K. {Ng}, E.~S. {Lo}, and R.~{Schober}, ``Robust beamforming for secure
  communication in systems with wireless information and power transfer,'' {\em
  IEEE Trans. Wireless Commun.}, vol.~13, pp.~4599--4615, Aug 2014.

\bibitem{control2016}
Y.~{Zeng}, R.~{Zhang}, and T.~J. {Lim}, ``Wireless communications with unmanned
  aerial vehicles: Opportunities and challenges,'' {\em IEEE Commun. Mag.},
  vol.~54, no.~5, pp.~36--42, 2016.

\bibitem{Tse2009Fundamentals}
D.~Tse and P.~Viswanath, {\em Fundamentals of wireless communication}.
\newblock Cambridge University Press, 2005.

\bibitem{zhu2019}
L.~{Zhu}, J.~{Zhang}, Z.~{Xiao}, X.~{Cao}, D.~O. {Wu}, and X.-G. Xia, ``3-{D}
  beamforming for flexible coverage in millimeter-wave {UAV} communications,''
  {\em IEEE Wireless. Commun. Lett.}, vol.~8, no.~3, pp.~837--840, 2019.

\bibitem{macro22020}
M.~{Di Renzo} {\em et~al.}, ``Reconfigurable intelligent surfaces vs. relaying:
  Differences, similarities, and performance comparison,'' {\em IEEE Open J.
  Commun. Soc.}, vol.~1, pp.~798--807, 2020.

\bibitem{yan2020}
W.~{Yan}, X.~{Yuan}, Z.~{He}, and X.~{Kuai}, ``Passive beamforming and
  information transfer design for reconfigurable intelligent surfaces aided
  multiuser {MIMO} systems,'' {\em IEEE J. Sel. Areas Commun.}, pp.~1--1, 2020.

\bibitem{he2020}
Z.~{He} and X.~{Yuan}, ``Cascaded channel estimation for large intelligent
  metasurface assisted massive {MIMO},'' {\em IEEE Wireless. Commun. Lett.},
  vol.~9, no.~2, pp.~210--214, 2020.

\bibitem{Mishra2019}
D.~{Mishra} and H.~{Johansson}, ``Channel estimation and low-complexity
  beamforming design for passive intelligent surface assisted miso wireless
  energy transfer,'' in {\em Proc. IEEE Int. Conf. Acoust., Speech Signal
  Process. (ICASSP)}, pp.~4659--4663, 2019.

\bibitem{liuhang2020}
H.~{Liu}, X.~{Yuan}, and Y.~A. {Zhang}, ``Matrix-calibration-based cascaded
  channel estimation for reconfigurable intelligent surface assisted multiuser
  {MIMO},'' {\em IEEE J. Sel. Areas Commun.}, pp.~1--1, 2020.

\bibitem{you2020}
C.~{You} and R.~{Zhang}, ``Hybrid offline-online design for {UAV}-enabled data
  harvesting in probabilistic {LoS} channels,'' {\em IEEE Trans. Wireless
  Commun.}, vol.~19, no.~6, pp.~3753--3768, 2020.

\end{thebibliography}

\end{document}